\documentclass[amsmath,amssymb,nofootinbib,prd]{revtex4}

\usepackage{graphicx}
\usepackage{epsfig}
\usepackage{rotate}
\usepackage{amsmath}
\usepackage{amssymb}
\usepackage{amsfonts}
\usepackage{enumerate}
\usepackage{afterpage}
\usepackage{color}

\def\DHLhksqrt#1#2{\setbox0=\hbox{$#1\sqrt{#2\,}$}\dimen0=\ht0
\advance\dimen0-0.2\ht0
\setbox2=\hbox{\vrule height\ht0 depth -\dimen0}%
{\box0\lower0.4pt\box2}}

\newcommand{\nngt}{$n_{\rm NG}$ }

\newcommand{\bk}{{\bf k}}
\newcommand{\bn}{{\bf n}}
\newcommand{\bV}{{\bf V}}
\newcommand{\De}{\Delta}
\newcommand{\de}{\delta}
\newcommand{\ka}{\kappa}
\newcommand{\Om}{\Omega}
\newcommand{\dd}{\partial}
\newcommand{\HH}{{\cal H}}

\definecolor{darkred}{RGB}{175,0,0}

\begin{document}

\title{Cosmological Measurements with General Relativistic Galaxy Correlations}

\author{Alvise Raccanelli$^{1,2,3}$, Francesco Montanari$^{4}$, Daniele Bertacca$^{5,6}$, Olivier Dor\'{e}$^{2,3}$, Ruth Durrer$^{4}$ \\~}

\affiliation{
$^1$ Department of Physics \& Astronomy, Johns Hopkins University, 3400 N. Charles St., Baltimore, MD 21218, USA \\
$^2$ Jet Propulsion Laboratory, California Institute of Technology, Pasadena CA 91109, USA \\
$^3$ California Institute of Technology, Pasadena CA 91125, USA \\
$^4$ D\'{e}partement de Physique Th\'{e}orique and Center for Astroparticle Physics, Universit\'{e} de
Gen\`{e}ve, 24 quai Ernest Ansermet, 1211 Gen\`{e}ve 4, Switzerland \\
$^{5}$ Argelander-Institut f\"ur Astronomie, Auf dem H\"ugel 71, D-53121 Bonn, Germany \\
$^{6}$ Physics Department, University of the Western Cape, Cape Town 7535, South Africa
}

\begin{abstract}
We investigate the cosmological dependence and the constraining power of large-scale galaxy correlations, including all redshift-distortions, wide-angle, lensing and gravitational potential effects on linear scales.
We analyze the cosmological information present in the lensing convergence and in the gravitational potential terms describing the so-called ``relativistic effects'', and we find that, while smaller than the information contained in intrinsic galaxy clustering, it is not negligible.
We investigate how neglecting them does bias cosmological measurements performed by future spectroscopic and photometric large-scale surveys such as SKA and Euclid.
We perform a Fisher analysis using the CLASS code, modified to include scale-dependent galaxy bias and redshift-dependent magnification and evolution bias.
Our results show that neglecting relativistic terms introduces an error in the forecasted precision in measuring cosmological parameters of the order of a few tens of percent, in particular when measuring the matter content of the Universe and primordial non-Gaussianity parameters.
Therefore, we argue that radial correlations and integrated relativistic terms need to be taken into account when forecasting the constraining power of future large-scale number counts of galaxy surveys.
\end{abstract}

\date{\today}

\maketitle


\section{Introduction}
At the beginning of the next decade, a variety of new galaxy surveys is expected to provide high precision data on galaxy clustering on very large scales.
Planned and proposed experiments such as the Prime Focus Spectrograph (PFS)~\cite{pfs},  DESI~\cite{desi}, SPHEREx~\cite{spherex}, the Square Kilometre Array (SKA)~\cite{ska}, Euclid~\cite{euclid} and WFIRST~\cite{wfirst} will probe cosmological volumes an order of magnitude larger than what has been provided by current surveys, so there is the need of a precise theoretical modeling of galaxy clustering.

Analyses of power spectrum and correlation function have been performed mainly using the so-called Kaiser formalism~\cite{Kaiser:1987, Hamilton:1997}, that assumes (other than linearity) the plane-parallel approximation, and it also sets the two galaxies of the pair at the same redshift. Classical analyses are also performed using Newtonian approximations, and the validity of this approach breaks down when probing scales approaching the horizon, where quantities have to be defined in a gauge-invariant way consistently including perturbations of observed redshift and volume.
Theoretical models for galaxy clustering on large scales were developed and tested including wide angle corrections to the Kaiser approximation and radial effects~\cite{Szalay:1997, Matsubara:1997, Matsubara:1999, Suto:1999, Matsubara:2000, Szapudi:2004, Papai:2008, Raccanelli:2010, Samushia:2012, Montanari:2012, Bonvin:2013}, and, more recently, including all  effects due to the fact that we observe galaxies on the physical, perturbed past light cone and not on some fictitious Friedman background~\cite{Yoo:2010, Challinor:2011, Bonvin:2011, Jeong:2012, Bertacca:2012, Yoo:2012, Raccanelli3D, Raccanelliradial, Lombriser:2013, DiDio:2013, DiDio:2013bqa, Yoo:2013, SKA:Camera}.

Future galaxy surveys are an excellent opportunity to advance our understanding of cosmology in general and Large Scale Structure (LSS) in particular, e.g. via constraining the Dark Energy equation of state or by testing General Relativity~\cite{Amendola:2012ys, Raccanelli:Growth}.
Furthermore, joint analysis with CMB experiments \cite{Planck:2015xua,Ade:2015ava} will bring better constraints on the primordial Universe.
Inflation models can be constrained studying, e.g., the primordial power spectrum and features induced by primordial non-Gaussianity.
The latter affect the galaxy correlation function on very large scales.
The interesting results of \cite{Matarrese:2000, Dalal:2008} show that primordial non-Gaussianity are detectable also via the two-point correlation function (or its Fourier transform, the power spectrum) since they enter as a scale-dependence of the large-scale linear galaxy bias.

The main goal of the present paper is to study the cosmological dependence of relativistic effects and the error in parameter estimation introduced when those terms are neglected.
We will also investigate which term is mainly responsible for the results obtained in~\cite{Camera:2014, Camera:2014b}, and we extend their previous analysis to other cosmological parameters.
We consider the SKA and Euclid surveys, consistently including galaxy bias, magnification bias and evolution bias.
We study the number counts angular power spectra at first order in perturbation theory and perform a Fisher matrix analysis to estimate cosmological parameters.
Our aim is not to forecast the most stringent and precise error bars for cosmological parameters, but to compare the standard Newtonian analysis to the fully relativistic one. For this we assume rather conservative specifications.
This yields relatively large error ellipses, but assures that our analysis is not affected by, e.g., the treatment of non-linear scales or by too optimistic survey specifications.
A detailed analysis of the goodness of the constraints per se is left as a future work including forecasts based on Markov chain Monte Carlo methods.

The paper is organized as follows.
In Section~\ref{sec:correlations} we introduce the formalism used throughout the paper and in Section~\ref{sec:surveys} we present the surveys specifications used to model number density as a function of redshift, the galaxy bias, the magnification, the evolution bias and discuss our Fisher analysis.
In Section~\ref{sec:cosmology} we investigate how relativistic effects can be used to constrain cosmological models and what is their effect on parameter estimation, and in Section~\ref{sec:cosdep} we show that relativistic effects contain cosmological information, and so they can in principle be used to constrain cosmological models.
Our results on how neglecting relativistic effects can modify the predicted cosmological measurements of future surveys are summarized in Section~\ref{sec:results}, and we discuss our conclusions in Section~\ref{sec:conclusions}.
Appendix \ref{sec:fNL_class} describes our implementation of primordial non-Gaussianity into the {\sc Class} code.


\section{Galaxy correlations in General Relativity}
\label{sec:correlations}
We cannot observe galaxies in real physical space at constant time. We observe them on our background light cone. Actually, for each galaxy we observe its direction $\bn$ in the sky as well as its redshift $z$. Taking into account that this redshift is perturbed by peculiar velocities is the well known ``redshift space distortion'' originally pointed out by Kaiser~\cite{Kaiser:1987}.
However, also the gravitational potential induces an apparent redshift modification and the direction into which we see the galaxy may not be the one into which the photons have been emitted. Taking into account all these relativistic effects to first order, one can define the {\it observed} galaxy over-density $\De_{\rm obs}(\bn,z)$ at fixed observed redshift and into a given observed direction $\bn$.  This is an observable and therefore gauge-invariant quantity which involves not only the density fluctuation and the velocity but also the Bardeen potentials $\Phi$ and $\Psi$. The fact that this expression is more complicated than the simple Kaiser formalism is not only a difficulty, but as we shall show, also an opportunity. It means that the number counts contain information not only on the density but also, e.g., about lensing, which is relevant and does affect cosmological parameter estimation.

\begin{figure*}[htb!]
\includegraphics[width=0.49\linewidth]{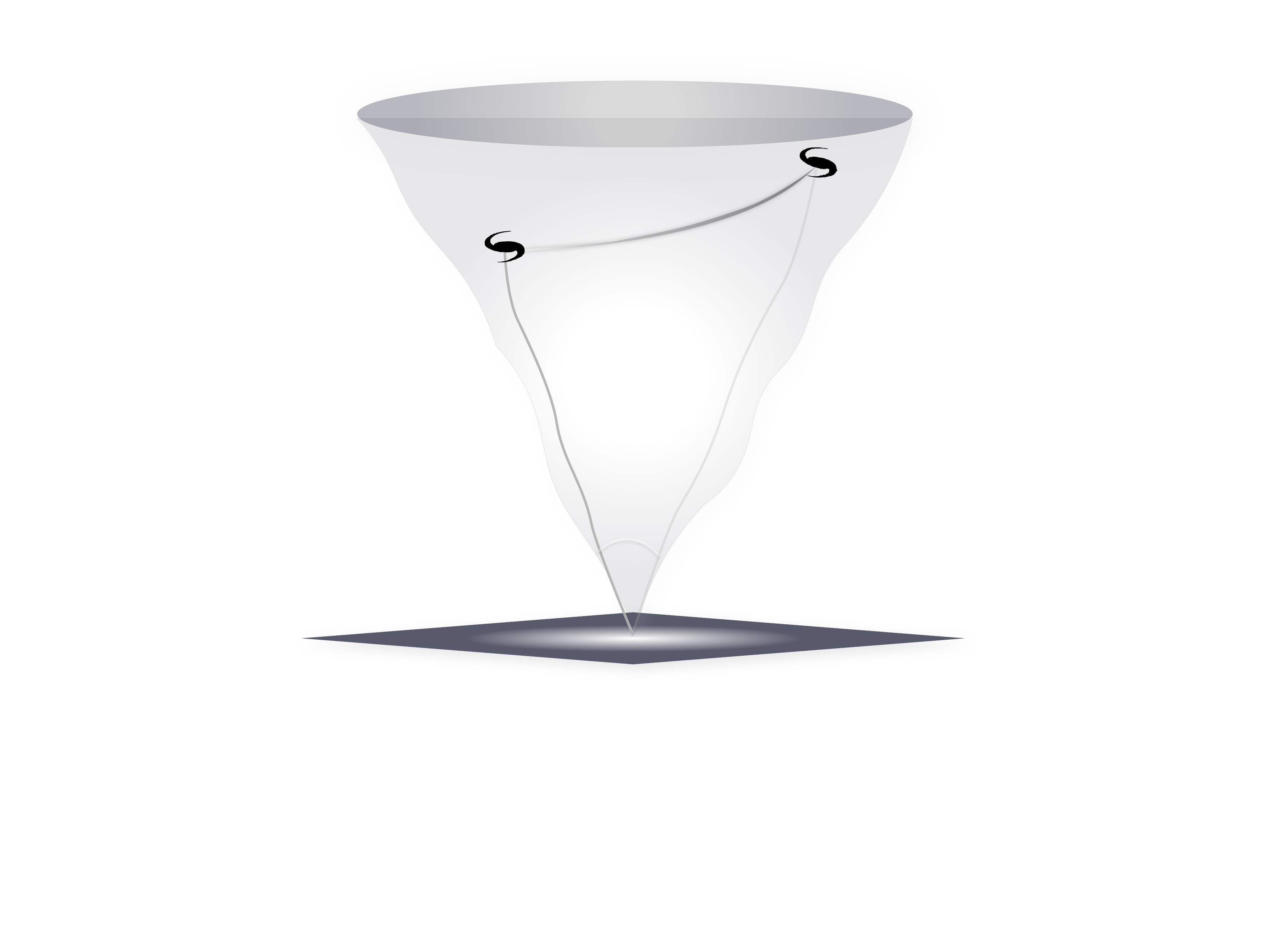}
\includegraphics[width=0.49\linewidth]{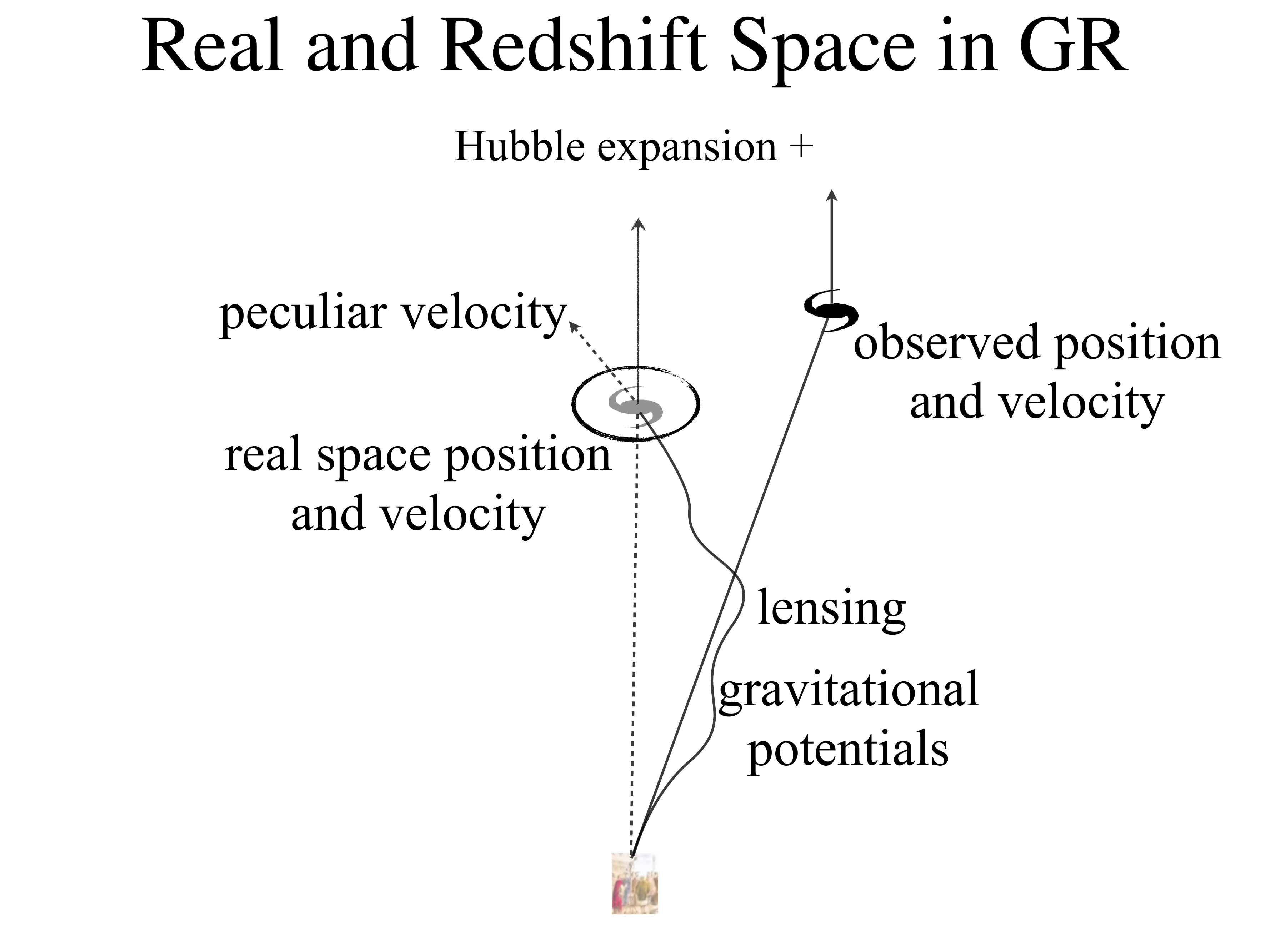}
\caption{{\it Left Panel:} We observe galaxy directions and redshifts on our perturbed past light cone.
{\it Right Panel:} Effect on the apparent position of a galaxy: peculiar velocity, gravitational potential and lensing modify the real position and velocity (in the center of the image, circled), to the observed one .}
\label{fig:geometry}
\end{figure*}

The total overdensity $\Delta_{\rm obs}(\bn,z)$ is not anymore just the {\it local} density field, but it is comprised also by {\it integrated} terms that correspond to lensing, time-delay and integrated Sachs-Wolf effects, due to the impact of cosmological perturbations on the path of photons. We can then express the total {\it observed} over-density as:
\begin{equation}
\label{eq:deltas}
\Delta_{\rm obs} ({\bf n}, z) = \Delta_{\delta}({\bf n}, z) + \Delta_{\rm rsd}({\bf n}, z) + \Delta_{\rm v}({\bf n}, z) + \Delta_{\rm \kappa}({\bf n}, z) + \Delta_{\rm pot}({\bf n}, z) \, ,
\end{equation}
where $\delta$ refers to the overdensity in the comoving gauge, $rsd$ and $v$ are peculiar veolcity (redshift space distortions in the Kaiser approximation) and Doppler effects,
$\kappa$ contains lensing convergence and $pot$ incorporates local and non local terms depending on Bardeen potentials and their temporal derivatives. The definition of these in terms of standard linear perturbation variables are given in Appendix~\ref{sec:fNL_class}. There we also present the transfer functions of these terms, including galaxy bias $b(z,k)$, magnification bias $s(z)$ (in the literature also the quantity $Q(z)=5s(z)/2$ is sometimes used) and evolution bias $f_{\rm evo}(z)$ (sometimes called $b_e(z)$).

In Figure~\ref{fig:geometry} we illustrate the apparent modifications to the position of a galaxy due to all the effects mentioned; in the ``standard'' case the only effect present is the one due to the Kaiser approximation for peculiar velocities that causes a spherical distribution in comoving coordinates space, to be observed as squashed along the line of sight (see e.g. Figure~1 of~\cite{Hamilton:1997}).
In the relativistic treatment there are also contributions from lensing convergence $\kappa$ and gravitational potential terms, that cause an apparent modification to the angular and radial position, respectively.
The convergence $\kappa$ is normally anti-correlated with the overdensity $\delta$ but the magnification bias which enters with the opposite sign can dominate it, especially at high redshift, see, e.g. Figure~12 of~\cite{Fosalba:2013mra}.
Because of this, a distribution of galaxies may appear more or less compact.
Finally, the gravitational potential terms directly involve Bardeen potentials, their time derivatives and their integrals along the line of sight (time-delay and integrated Sachs-Wolf effects).

These correlations have been investigated recently in~\cite{Bertacca:2012, Raccanelliradial, DiDio:2013, Bonvin:2013}, and it has been shown that, for radial correlations with a large $\Delta \, z$, relativistic terms can be a significant contribution to the correlation~\cite{Raccanelliradial}.

Since $\De_{\rm obs}(\bn,z)$ is a function of direction and redshift, its power spectra are defined for two redshifts $z$ and $z'$,  $C_\ell(z,z')$, see~\cite{Bonvin:2011}. In Appendix~\ref{sec:fNL_class} we define the transfer function $\Delta_{\ell}^{W_i}(k)$ of the number counts given in Equation~(\ref{eq:deltas}). This computes the contribution to  $\De_{\rm obs}$ from wave number $k$ estimated in a bin with mean redshift $z_i$ and window function $W_i$. From the transfer function, the angular power spectra are determined by integration over the primordial curvature spectrum~\cite{DiDio:2013}:
\begin{equation}
\label{eq:Cls}
C_{\ell}^{ij} = 4\pi \int \frac{dk}{k} \Delta_{\ell}^{W_i}(k) \Delta_{\ell}^{W_j}(k) \mathcal{P}_{\mathcal{R}}(k) \;,
\end{equation}
where $\mathcal{P}_{\mathcal{R}}(k)$ is the primordial power spectrum of curvature perturbations. The superscript $W_i$ indicates that these transfer already include an integral over the window function $W_i$.
More details are given in Appendix~\ref{sec:fNL_class}. 

We note that in the standard analysis of power spectra $P(k)$ it is customary to use only auto-correlations within a certain redshift bin of a galaxy survey, and not the cross-bin correlations as redshift bins are treated as independent.
In the following we show that it is important to be able to model consistently cross-bin correlations, since they are dominated by the non-negligible lensing convergence $\kappa$. This is very interesting since  $\kappa$  is the (spherical) Laplacian of the lensing potential $\psi$, hence their spectra are related by 
$C^\kappa_\ell = [\ell(\ell+1)]^2C^\psi_\ell$. Therefore, if galaxy surveys are sensitive to $\kappa$, we can in principle use them to determine the lensing potential, usually inferred by shear measurements.


\section{Future galaxy surveys}
\label{sec:surveys}
To investigate the observational consequences of including relativistic corrections we consider two examples of future surveys: the HI galaxy survey planned with the SKA~\cite{skaHI} as an example of a spectroscopic survey, and the Euclid photometric survey~\cite{euclid, Amendola:2012ys}.
We use specifications for SKA which are consistent with the $5\mu$Jy cut of \cite{Santos:2015hra}, and the photometric specifications for the Euclid survey.
In Figure~\ref{fig:nzbz}, ~\ref{fig:sz} we plot the predicted number of galaxies, galaxy bias and magnification bias that we will use for our forecasts.

\subsection{Survey specifications}
\subsubsection{Galaxy distribution}
\begin{figure*}[htb!]
\includegraphics[width=0.49\linewidth]{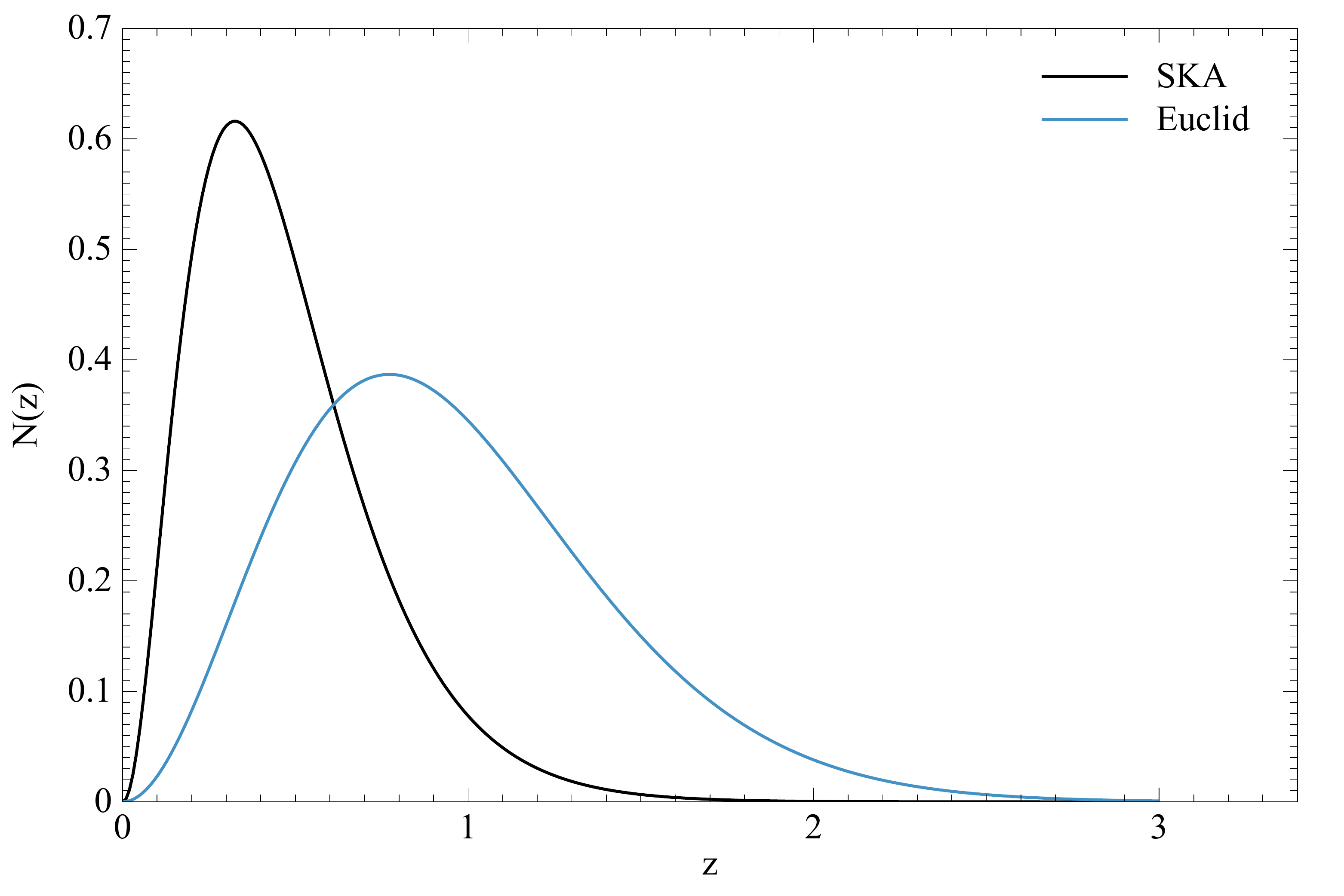}
\includegraphics[width=0.49\linewidth]{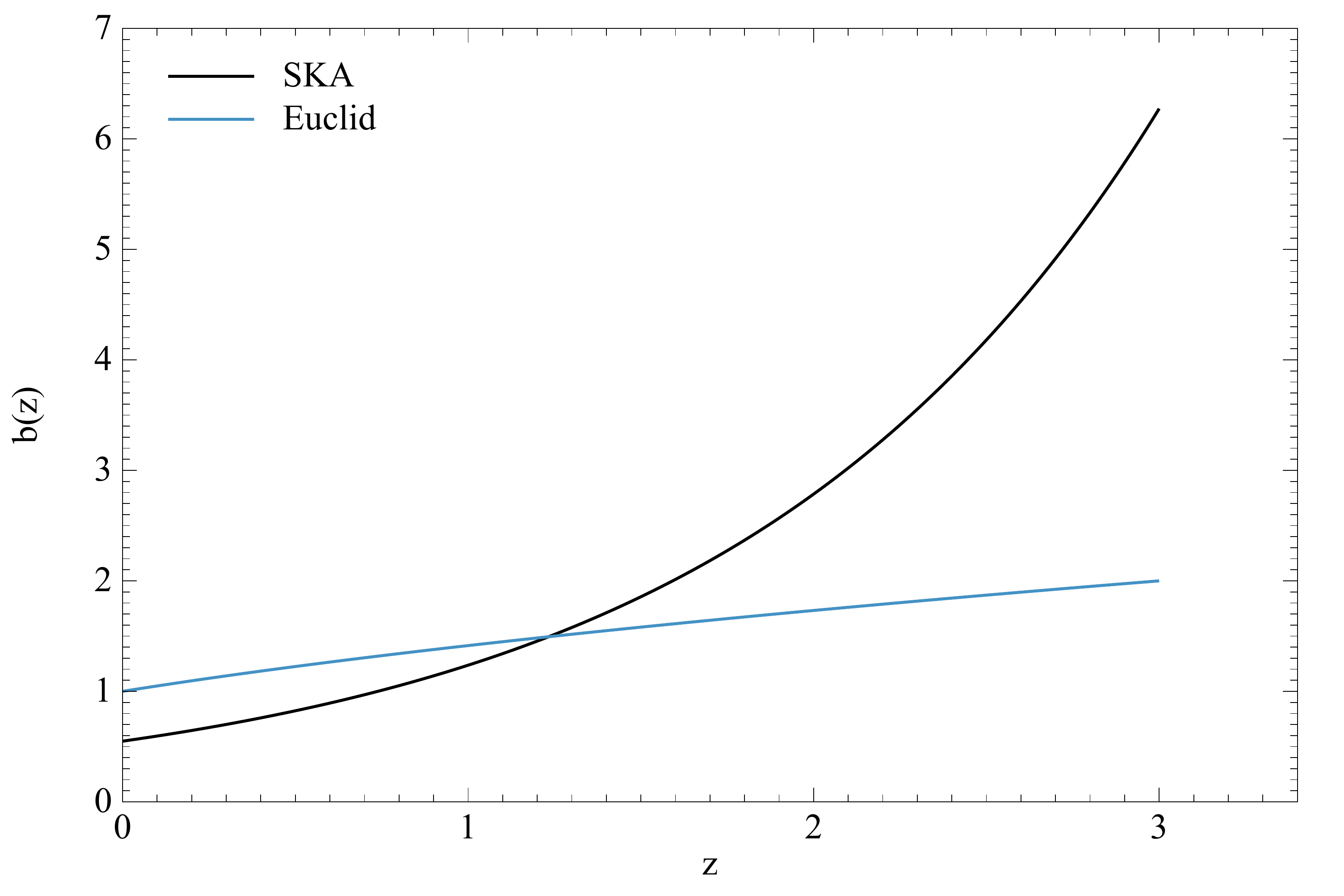}
\caption{
Redshift distribution, normalized to unity, (left panel) and bias (right panel) for the SKA and Euclid surveys considered in this paper.
}
\label{fig:nzbz}
\end{figure*}

The number of galaxies per steradian and per unit redshift for SKA \cite{Santos:2015hra} is given by:
\begin{equation}
\frac{dN}{dzd\Omega} = \left(\frac{180}{\pi}\right)^2 10^{c_1} z^{c_2} e^{- c_3 z} \, ;
\end{equation}
we use $c_1=6.7767$, $c_2=2.1757$, $c_3=6.6874$; the survey covers $30,000{\rm\ deg}^2$, i.e. $f_{\rm sky}=0.73$.
We divide the survey into 5 bins of equal depth $\Delta z=0.39$ in the range $0.05<z<2$.

For the Euclid photometric survey we use:
\begin{equation}
\frac{dN}{dzd\Omega} = 3.5\times10^8 z^2 \exp\left[-\left( \frac{z}{z_0} \right)^{3/2}\right] \, , \,\,\, 0<z<2 \, ,
\end{equation}
where $z_0=z_{\rm mean}/1.412$ and the median redshift is $z_{\rm mean}=0.9$; $f_{\rm sky}=0.375$. This number density is consistent with the official specification for the density of galaxies, $d=30 \, {\rm arcmin}^{-2}$.

In both cases, these specifications will give a total number of sources detected of $\sim 10^9$.

\subsubsection{Galaxy bias}
The galaxy bias for SKA is approximated as \cite{Santos:2015hra}:
\begin{equation}
b_G(z) = c_4 e^{c_5 z} \, ,
\end{equation}
with $c_4=0.5887$, $c_5=0.8137$.

As for  Euclid, we model the galaxy bias as:
\begin{equation}
b(z)=\sqrt{1+z} \; .
\end{equation}

These fits are conveniently taken into account in the implementation of primordial non-Gaussianities discussed in Appendix~(\ref{sec:fNL_class}).

\subsubsection{Evolution bias}
The \emph{rsd} and \emph{pot} terms also include evolution bias, $f_{\rm evo}(z)$, (see e.g.~\cite{Jeong:2012, Bertacca:2012, DiDio:2013bqa}). This is due to the fact that new galaxies form so that the true number density of galaxies does not simply scale like $a^{-3}$ where $a$ is the cosmological scale factor. The evolution bias is given by:
\begin{equation}
\label{eq:fevo}
f_{\rm evo}(z) = \frac{1}{aH}\frac{d}{d\tau}\ln\left( a^3 \frac{d\bar{N}(z,L>L_{\rm lim})}{dzd\Omega} \right) \;,
\end{equation}
where $\tau$ is the conformal time and $H$ is the Hubble parameter.
Here $d\bar{N}(z,L>L_{\rm lim})/dz/d\Omega$ indicates the true number density of galaxies (not necessarily observed) per redshift and per solid angle present in the Universe above a certain threshold $L_{\rm lim}$ in luminosity.
It can be estimated from the luminosity function \cite{Challinor:2011}.
However, given the uncertainties in the modelling of galaxy evolution , for simplicity as in \cite{Camera:2014} we assume that the observed $dN(z)/dz/d\Omega$ in Figure \ref{fig:nzbz} still gives a good approximation to estimate $f_{\rm evo}(z)$.
This comes from the fact that the evolution bias only appears in subleading terms (wide-angle velocity and \emph{pot}) so that uncertainties in its modeling do not  significantly affect our results.

\subsubsection{Magnification bias}
For SKA we interpolate the results of \cite{Camera:2014} to $5\mu$Jy:
\begin{equation}
s(z) = c_6 + c_7 e^{-c_8 z}
\end{equation}
with $c_6=0.9329$, $c_7=-1.5621$, $c_8=2.4377$, and use a constant $s(z_i)$ within each bin centered at $z_i$.

\begin{figure*}[htb!]
\includegraphics[width=0.49\linewidth]{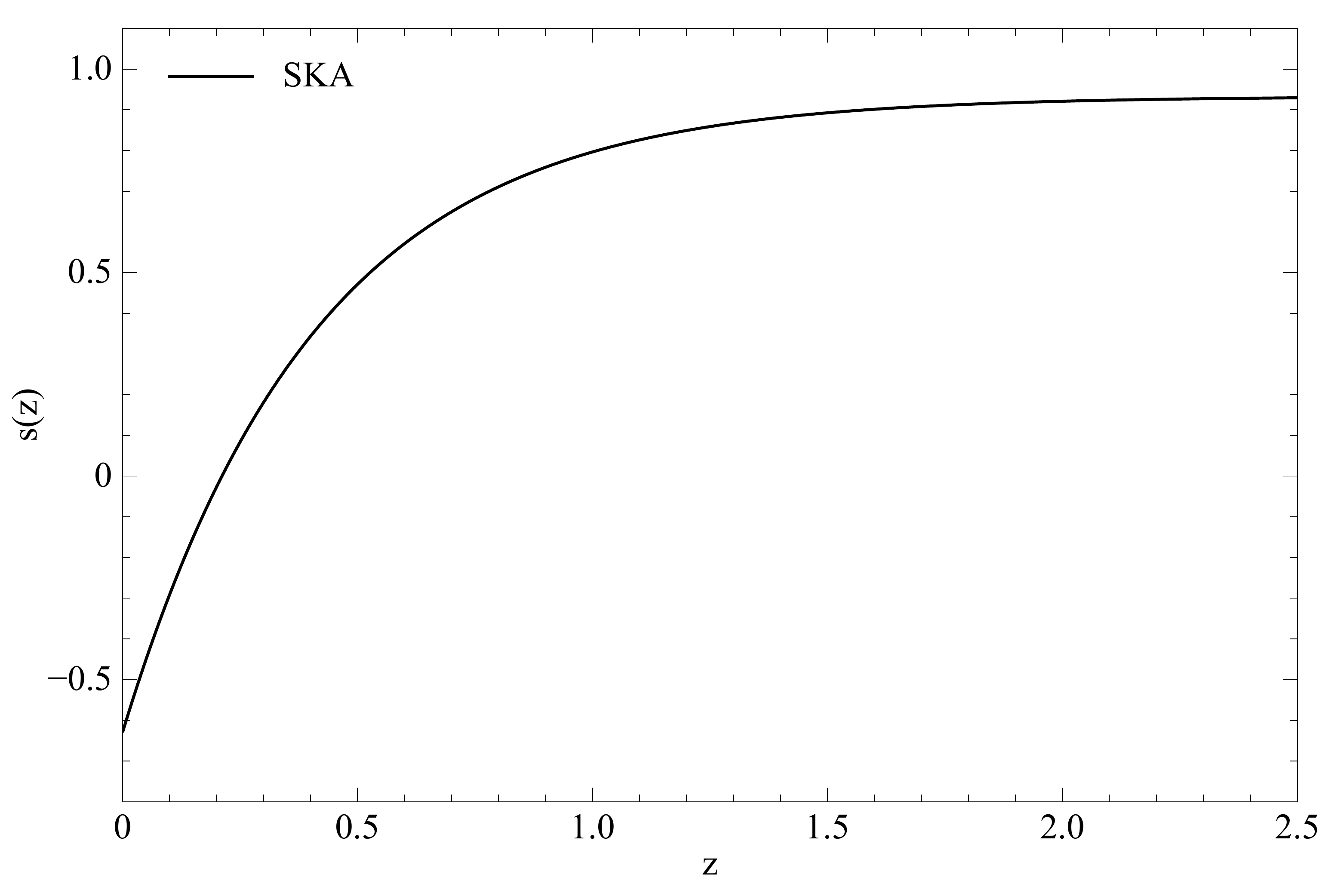}
\includegraphics[width=0.49\linewidth]{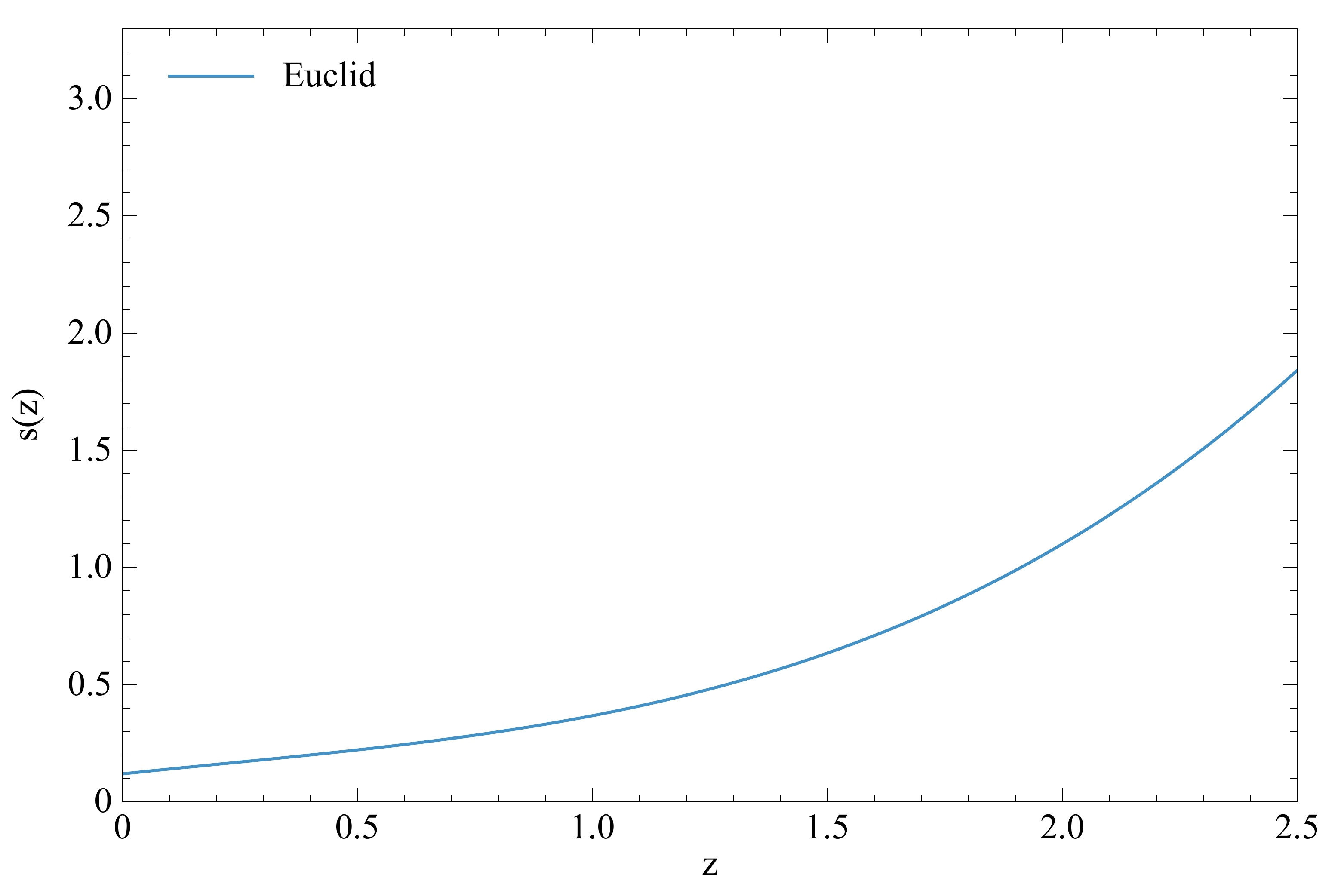}
\caption{
Magnification bias for the SKA (left panel) and Euclid (right panel) surveys considered in this paper.
}
\label{fig:sz}
\end{figure*}

We use the magnification bias for the photometric Euclid survey computed in~\cite{Montanariprep}, given by the fit:
\begin{equation}
s(z)=s_0 + s_1 z + s_2 z^2 + s_3 z^3 \, 
\end{equation}
with $s_0=0.1194$, $s_1=0.2122$, $s_2=-0.0671$ and $s_3=0.1031$. This result is consistent also with other analyses, e.g.~\cite{Liu:2013}.
In Figure~\ref{fig:sz} we plot the magnification bias used in this work.
We modified the {\sc CLASS} code to introduce a magnification bias $s(z_i)$ depending on the central redshift $z_i$ of each bin.

Cosmic magnification changes the relative number of sources detected at a given redshift and fixed magnitude limit, and in this way lensing can modify the effect of cosmological parameter variations on clustering (see e.g.~\cite{Matsubara:2000, Liu:2013}):
\begin{equation}
\label{eq:cosmag}
n_{\rm obs}(z) = n_g(z) [1 + \de+ (5s(z)-2) \kappa] \, ,
\end{equation}
where $n_{\rm obs}, n_g$ are the observed and intrinsic number of sources, respectively, $s$ is the magnification bias and $\kappa$ is the convergence.
We stress that this effect modifies the effect of primordial non-Gaussianity also, due to changes in the relative redshift distribution of observed sources, and so lensing effects are important for measurements of $f_{\rm NL}$, a fact that is not often taken into consideration.

\subsection{Fisher matrix analysis}
\label{sec:fisher}
In order to predict the constraining power of a survey one often uses a Fisher matrix analysis~\cite{Fisher:1935, tegmark97}. The Fisher matrix, $F_{\alpha\beta}$, indicates how well a given set of cosmological parameters $(\vartheta_\alpha)$  under consideration can be measured. The bigger the Fisher matrix elements, the better the parameters can be determined by the given survey. If the errors (in the parameters) would be Gaussian, the ellipsoid in parameter space given by:
$$ \left\{(\vartheta_{\alpha})_{\alpha=1}^n \left|   \sum_{\alpha\beta=1}^n F_{\alpha\beta}(\vartheta_\alpha-\bar\vartheta_\alpha)(\vartheta_\beta -\bar\vartheta_\beta) \leq \Delta_n\chi^2  \right. \right\}\, ,$$
where the Inverse Regularized Gamma function gives the difference to the minimum $\chi^2$ as $\Delta_1\chi^2=1$ and $\Delta_2\chi^2=2.30$ for $n=1,2$ degrees of freedom, respectively. This determines the 1$\sigma$ error bars or 68\% confidence region of parameter space around some fiducial values $(\bar\vartheta_{\alpha})$. In reality errors are not Gaussian but the Fisher ellipses usually still give a reasonable indication of the precision with which a given parameter can be determined. In our situation, where we measure the $C_\ell^{ij}$ with a certain precision $\sigma_{C_\ell}$, the Fisher matrix is given by:
\begin{equation}
F_{\alpha\beta} =
\sum_{(ij)(pq)}  \sum_{\ell} \frac{\partial C_\ell^{ij}}{\partial \vartheta_\alpha}
\frac{\partial C_\ell^{pq}}{\partial
\vartheta_\beta} {\sigma_{C_\ell [(ij),(pq)]}^{-2}} \, , 
\label{eq:Fisher}
\end{equation}
where the derivative is evaluated at fiducial values $\bar \vartheta_{\alpha}$ and $\sigma_{C_\ell}$ are errors in the power spectra. For Gaussian fluctuations they can be determined by Wick's theorem and are given by (see e.g.~\cite{DiDio:2013}):
\begin{equation}
\label{eq:err-clgt}
\sigma^2_{C_{\ell \, \rm[(ij), (pq)]}} = \frac{\tilde{C}_{\ell}^{\rm (ip)} \tilde{C}_{\ell}^{\rm (jq)} + \tilde{C}_{\ell}^{\rm (iq)} \tilde{C}_{\ell}^{\rm (jp)}}{(2\ell+1)f_{\rm sky}} \, ,
\end{equation}
where $\tilde{C}_{\ell} $ denote the observed correlation multipoles which include shot noise errors:
\begin{equation}
\tilde{C}_{\ell}  = C_{\ell}^{ij} + \frac{\delta_{ij}}{dN(z_i)/d\Omega} \, ,
\end{equation}
and $dN(z_i)/d\Omega$ is the average number of sources per steradian within the bin $z_i$.
Even with primordial non Gaussianity this is a good approximation for the error in the power spectra.
Note that we sum over the matrix indices $(ij)$ with $i\leq j$ and $(pq)$ with
$p \leq q$ which run from 1 to the number of bins.

To perform the Fisher analysis, we parameterize our cosmology using the following parameters:
\begin{equation}
\label{eq:paratriz} 
{\bf P} \equiv \{ w_{\rm 0}, w_{\rm a}, n_{\rm s}, \alpha_{\rm s}, \Omega_{\rm cdm}, \Omega_{\rm b}, h, f_{\rm NL}, n_{\rm NG} \} \,.
\end{equation}
We have set spatial curvature  $K=0$; $h$ parameterizes the present Hubble parameter, $H_0 = h 100$km/s/Mpc;
$\{w_0,w_a\}$ parameterize the dark energy equation of state, $w= w_0 +w_a(1-a)$; $\{n_{\rm s}, \alpha_{\rm s} \}$ are the primordial power spectral index and its running (see definition in the next section); $\{\Omega_{\rm cdm}, \Omega_{\rm b} \}$ are the density parameters of cold dark matter and of baryons respectively; $\{f_{\rm NL}, n_{\rm NG} \}$ are the primordial non-Gaussianity parameter and its spectral tilt (see definition in the next section).
We choose the fiducial values $\tilde w_0=-1$, $\tilde w_a=0.05$, $\tilde\alpha_s=0.01$, $\tilde n_{NG}=0.1$, $\tilde f_{\rm NL}=1$, $\tilde h= 0.67, \tilde\Om_{\rm cdm}= 0.25$ and $\tilde\Om_{\rm b}= 0.05$.

In doing a real data analysis, one should use the range $2 < \ell < k_{\rm max} \chi(z) $, but this requires involved algorithms to carefully exclude non-linear scales \cite{DiDio:2013} and to compute the beginning of the quasi-linear regime when cross-correlating different redshift bins.
Here we are mostly interested in a first assessment of the importance of relativistic effects as compared to the standard analysis, so to simplify our study, we assume a  conservative $\ell_{\rm max}=$200 at all redshifts.
Even if some of the relativistic effects are more sensitive to low multipoles, such a conservative cut on non-linear scale does not bias our constraints towards an ad-hoc enhancement of relativistic effects compared the Newtonian case. In fact some of the most relevant terms (e.g. lensing) also affect high multipoles, see Figures~\ref{fig:Cls_ska},~\ref{fig:Cls_euclid}.


\section{Estimating Cosmological parameters with relativistic galaxy correlations}
\label{sec:cosmology}
The main goal of this paper is to investigate the impact of relativistic corrections on galaxy clustering analyses and their constraining power by looking at the predicted $C_{\ell}(z,z')$ from future galaxy surveys.
We focus on constraints on dark energy and on the parameters defining the initial perturbations.
Note that we are studying the impact of relativistic terms relative to the Newtonian ones and we do not perform a forecast of survey capabilities; a detailed analysis of survey specifications and observational strategies together with an MCMC analysis including the nuisance parameters of the survey is required to correctly assess the proper forecasted measurements for specific instruments. This is not the focus of this work.

In this paper we compare constraints using a traditional analysis (but including wide-angle and local relativistic corrections; see~\cite{Raccanelliradial} for more details) with a fully relativistic one. We want to study how much information is present in the relativistic terms which have been neglected in the past. In order to do this, we compare two Fisher matrix analyses: 
 the Newtonian case, referred to as ``nwt'' where we include intrinsic clustering plus peculiar velocity effects (in the formalism of Equation~\ref{eq:deltas}, this corresponds to the first two terms), using only auto-correlation functions within the redshift bins used,  and the fully relativistic case, referred to as ``rel''  where we include all the terms and also cross-bin correlations.

\begin{figure*}[htb!]
\includegraphics[width=0.49\linewidth]{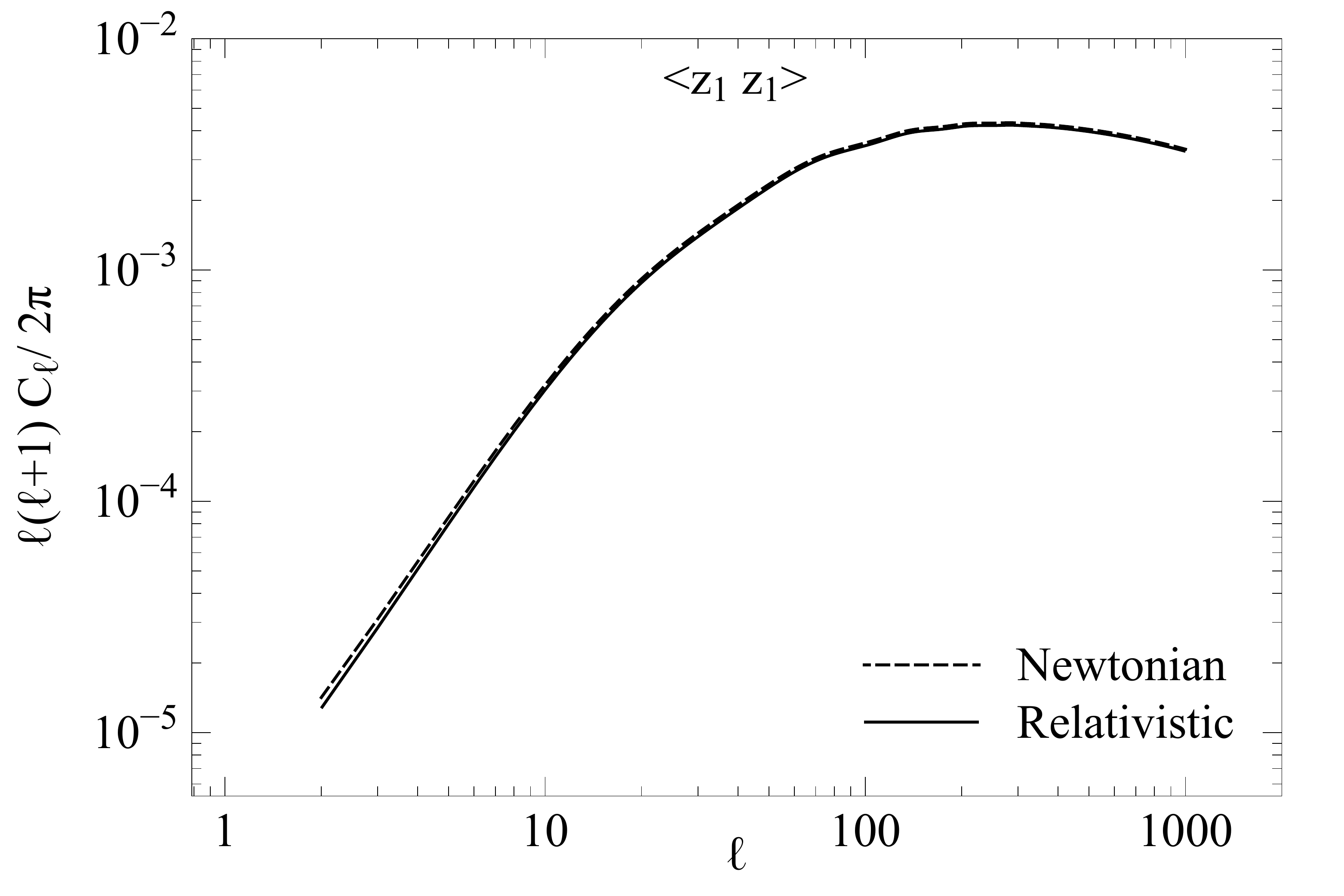}
\includegraphics[width=0.49\linewidth]{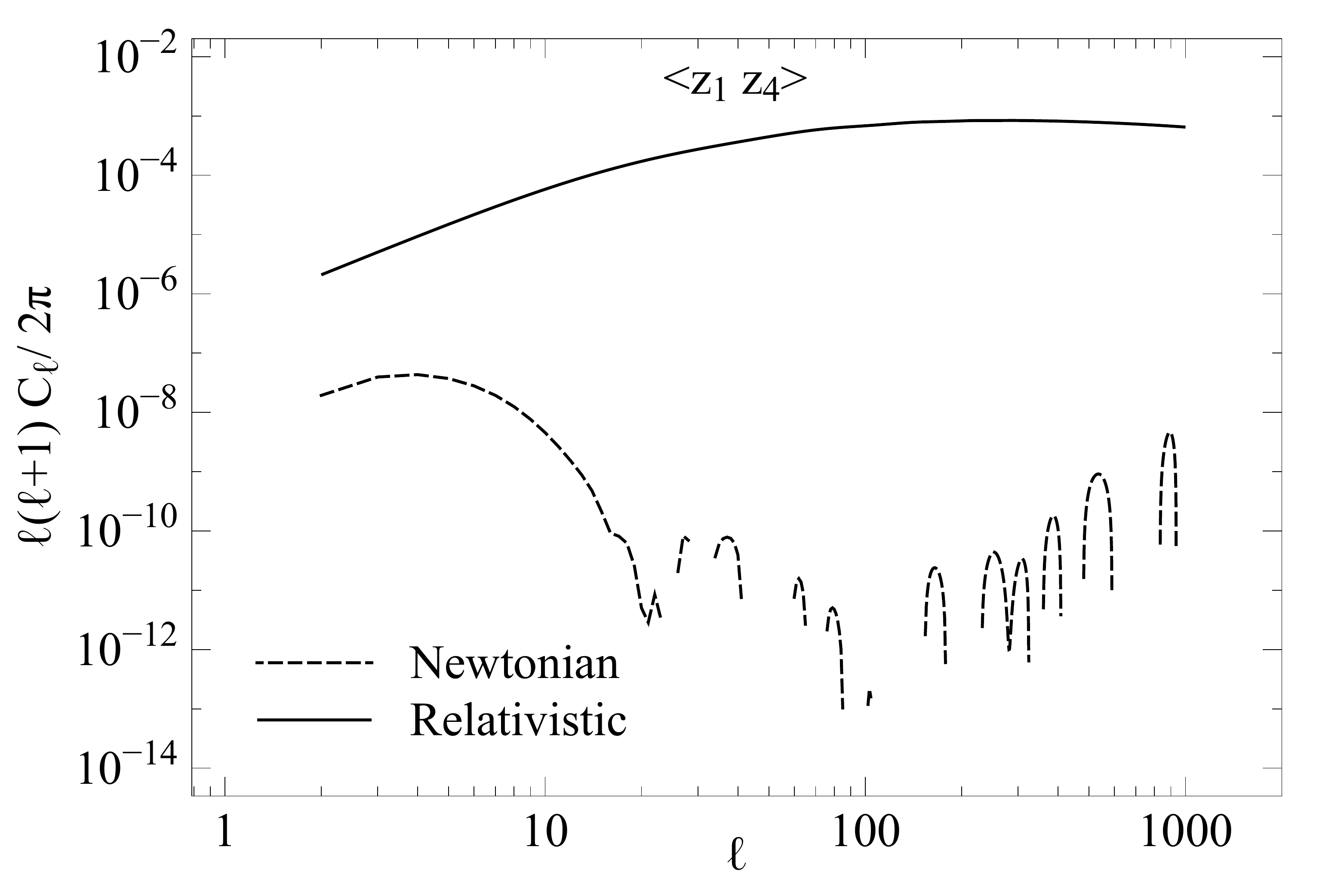}
\includegraphics[width=0.49\linewidth]{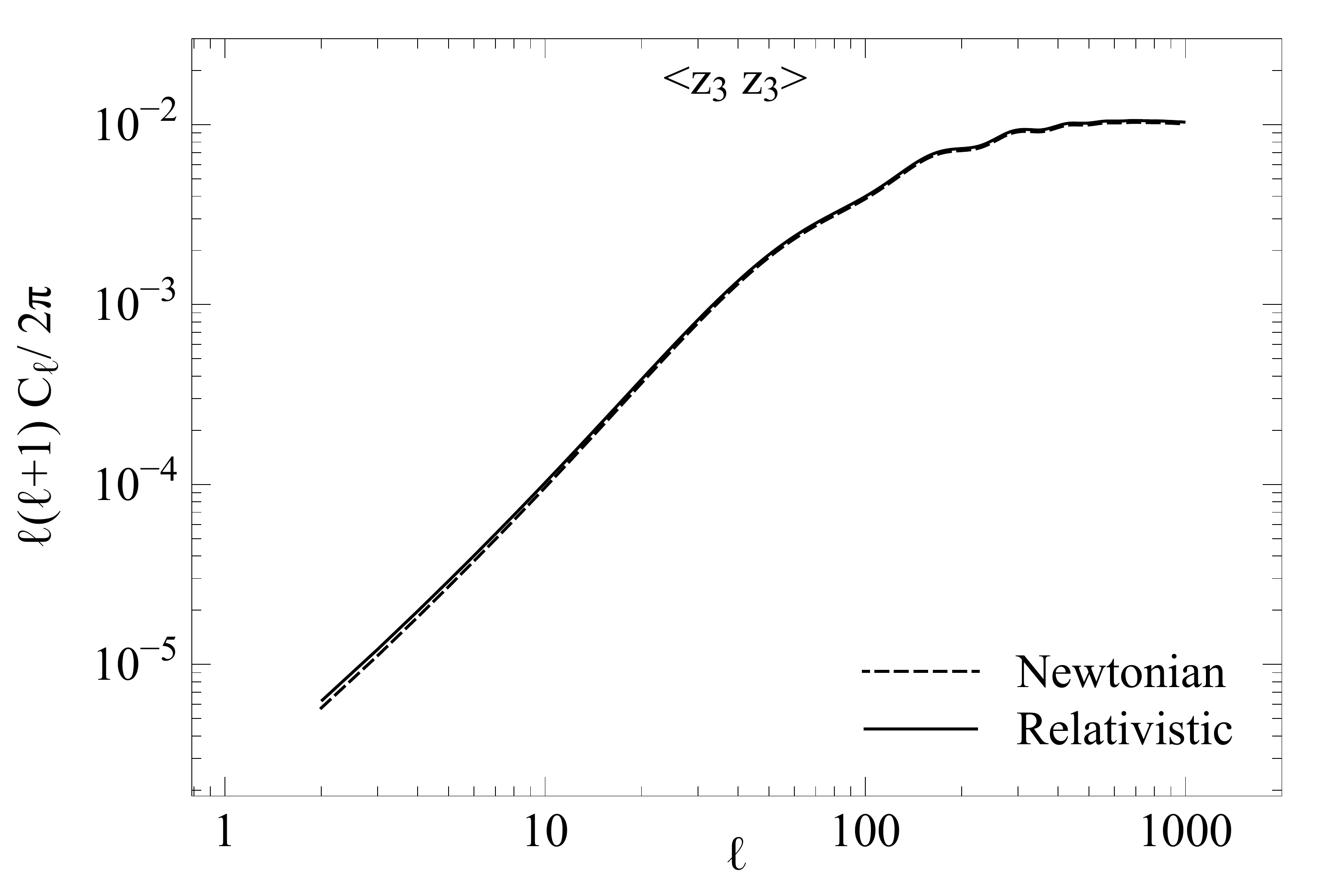}
\includegraphics[width=0.49\linewidth]{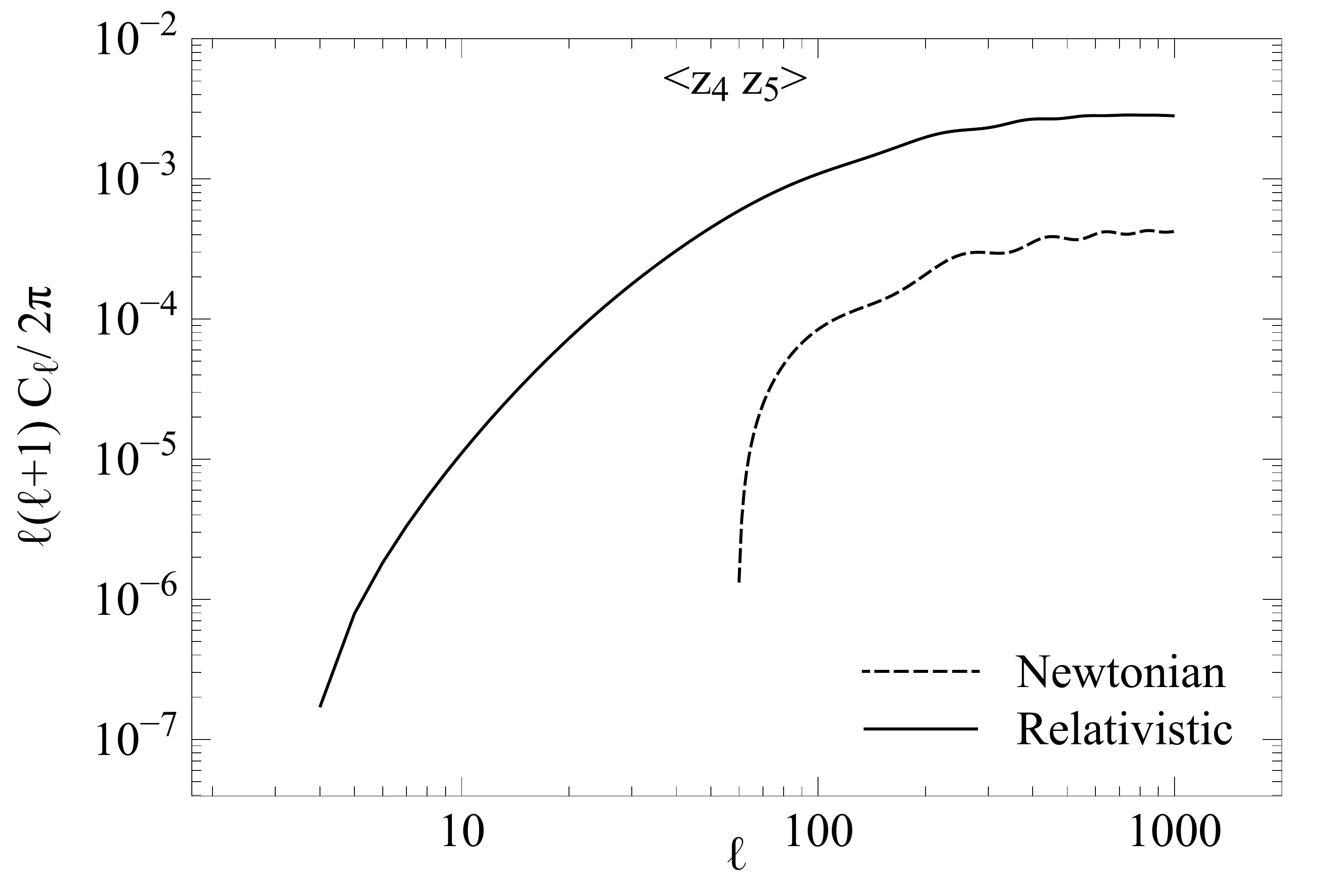}
\caption{
Power spectra in the Newtonian and Relativistic cases for different $C_{\ell}(z_A,z_B)$ for the spectroscopic SKA galaxy survey. Differences in the auto-correlations are small and at very low $\ell$, while cross-bin correlations are dominated by integrated relativistic terms.
}
\label{fig:Cls_ska}
\end{figure*}

\begin{figure*}[htb!]
\includegraphics[width=0.49\linewidth]{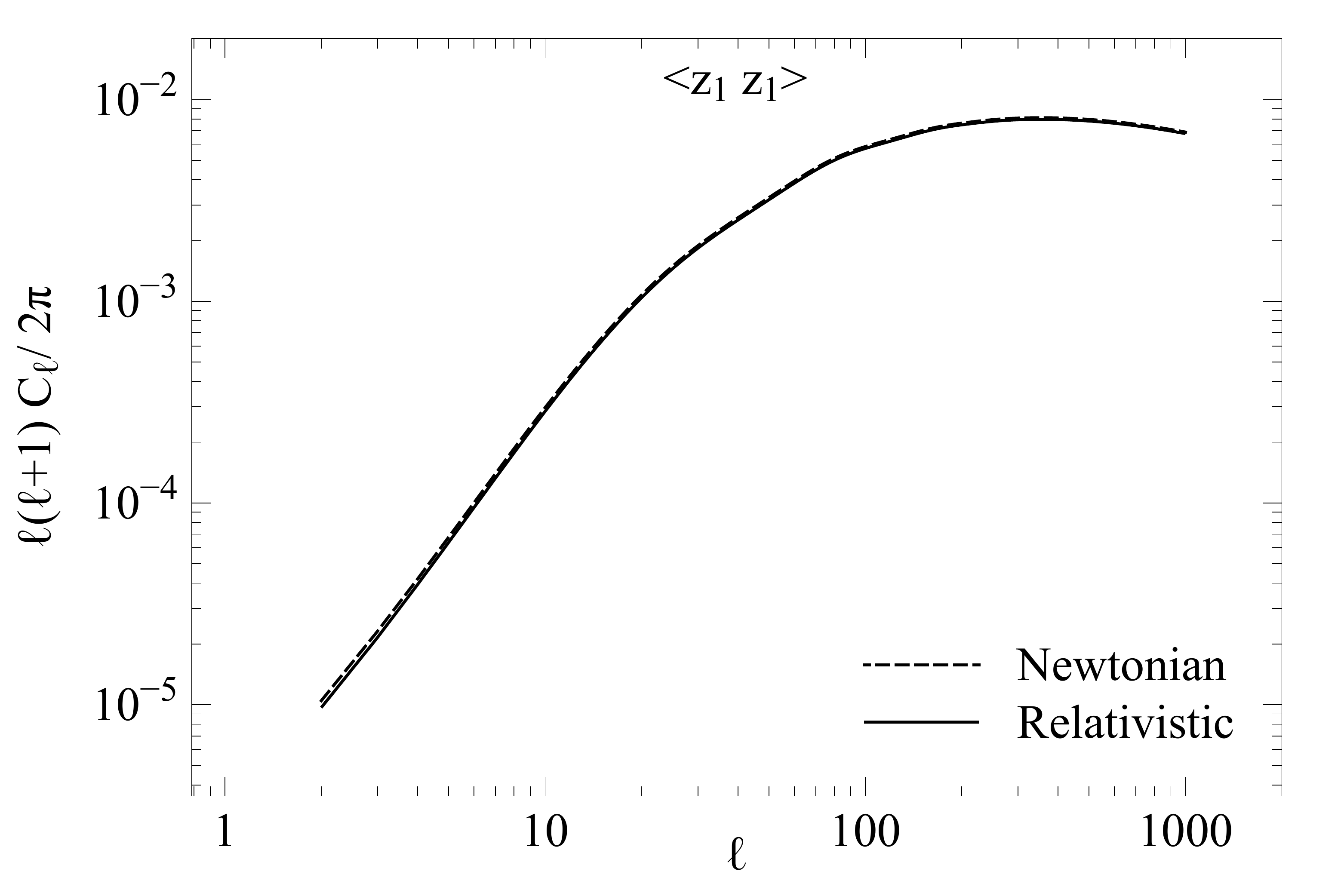}
\includegraphics[width=0.49\linewidth]{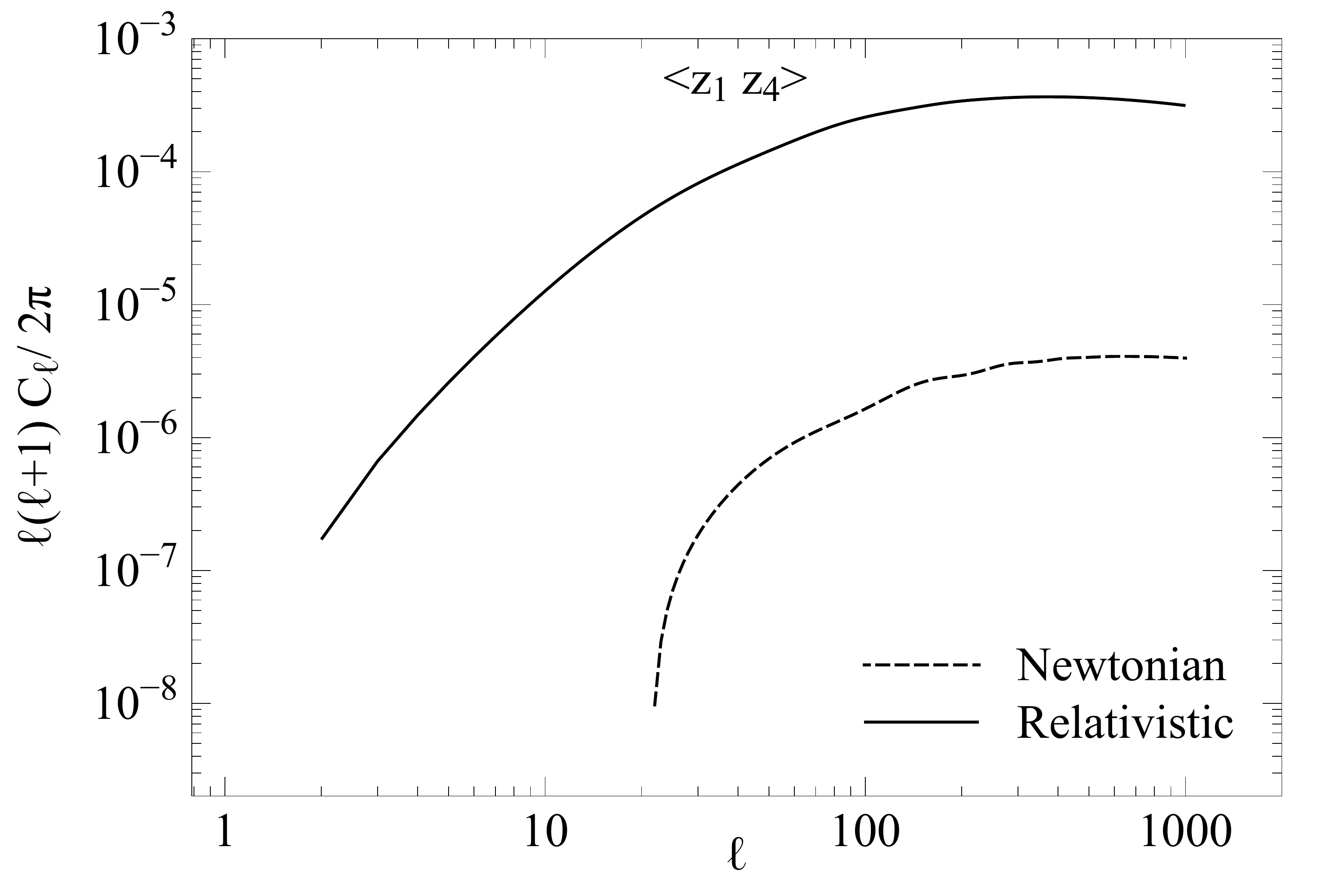}
\includegraphics[width=0.49\linewidth]{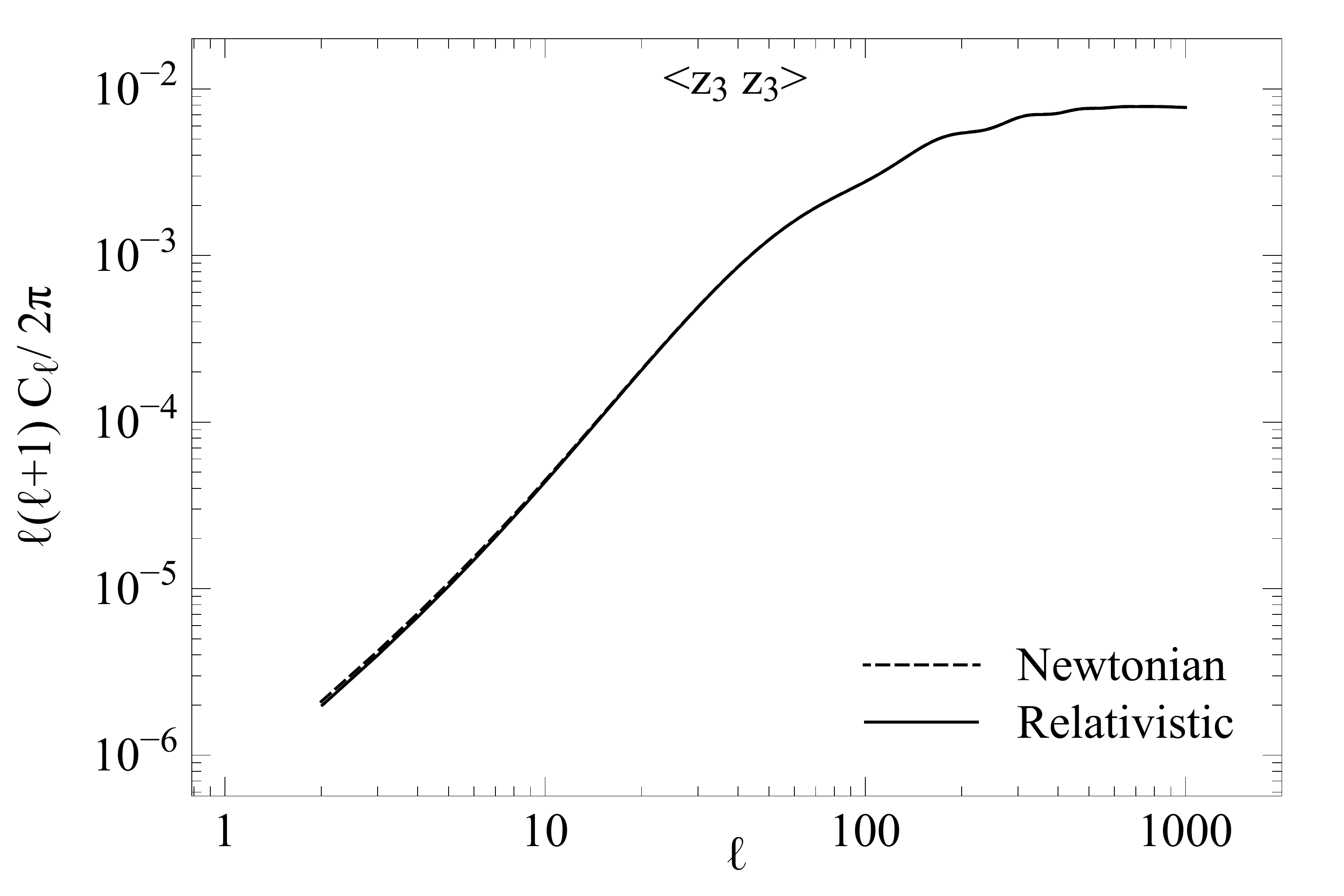}
\includegraphics[width=0.49\linewidth]{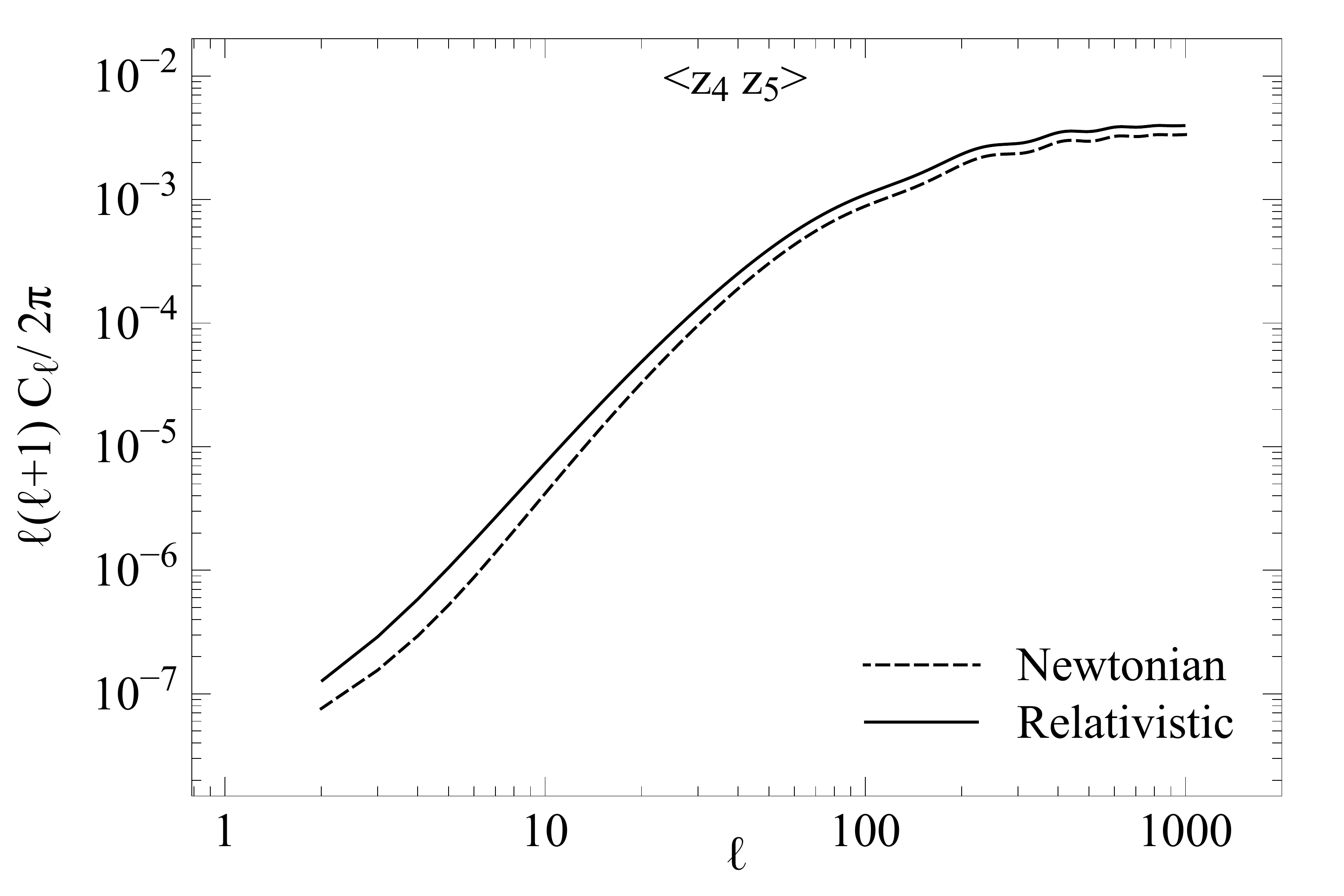}
\caption{
Power spectra in the Newtonian and Relativistic cases for different $C_{\ell}(z_A,z_B)$ for the photometric Euclid survey. Differences in the auto-correlations are small and at very low $\ell$, while cross-bin correlations are dominated by integrated relativistic terms.
}
\label{fig:Cls_euclid}
\end{figure*}

In Figure~\ref{fig:Cls_ska},~\ref{fig:Cls_euclid} we compare the angular power spectra in the Newtonian and relativistic case for the SKA (spectroscopic) and Euclid (photometric), respectively. In the plots we refer to $<z_A z_B>$ to indicate the correlations $C_{\ell}(z_A,z_B)$. In the case of spectroscopic surveys we assume top hat bins, while we use Gaussian bins in the case of photometric surveys; more details on the modeling and computation of $C_{\ell}(z_A,z_B)$ are given in Appendix~\ref{sec:fNL_class}.
In both cases, small differences are visible in the auto-correlations, with larger differences at small $\ell$; on the cross-bin correlations the differences are very large, because these correlations are dominated by the lensing convergence term.
In particular, non adjacent bin cross-correlations are nearly zero in the newtonian case for spectroscopic surveys, as expected. 
We note that oscillations in e.g. the $\langle z_1 z_4 \rangle$ case plotted here are not physical but due to numerical precision for such low (negligible) amplitudes.
This justifies the standard assumption that in spectroscopic surveys the cross-bin correlations contain virtually no cosmological information, but, as firstly noted in~\cite{Raccanelliradial}, we stress the fact that in the relativistic case, gravitational potential and lensing convergence effects introduce radial cross-bin large correlations.
Nearest cross-bin correlations, e.g. $\langle z_4 z_5 \rangle$ from Euclid are in general larger compared to the SKA ones in the Newtonian case, because photometric redshift uncertainties assumed for Euclid cause an effective cross-bin overlap (see e.g.~\cite{Blake:2005}). This means that local Newtonian term are still non-negligible compared to the integrated relativistic ones.
We will investigate the cosmological information present in these correlations in the rest of this Section.

\subsection{Dynamical Dark Energy}
\label{sec:de}
The first step towards understanding the nature of dark energy is to clarify whether it is a simple cosmological constant or whether it originates from a dynamical degree of freedom, i.e. it varies with time. 
Dynamical dark energy can be distinguished from the cosmological constant by considering the evolution of the equation of state of dark energy, $w = \frac{p}{\varrho}$,
where $p$ and $\varrho$ are the pressure and energy density of the ``dark energy fluid'', respectively. For a cosmological constant model, $w=-1$, while for dynamical models $w=w(a)$, where $a$ is normalized to unity today, $a_0=1$.

To evaluate the potential to constrain the dynamics of different models of dark energy with galaxy surveys, we adopt the following simple phenomenological parameterization for the dark energy equation-of-state (EoS) $w$~\cite{linder03, linder05}:
\begin{align}
w(a) = w_0 + w_a (1-a) \, .
\end{align}

\begin{figure*}[htb!]
\includegraphics[width=0.47\linewidth]{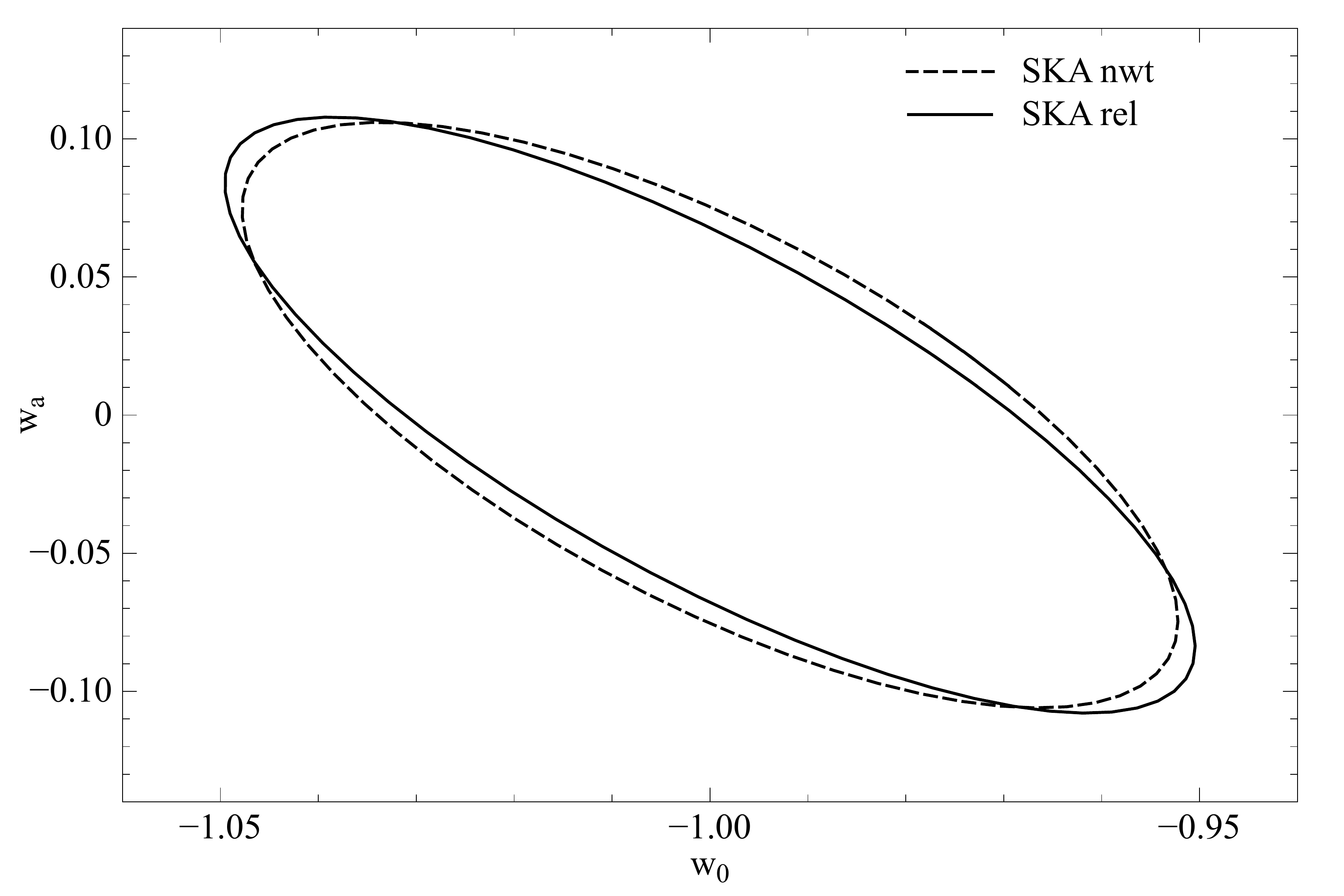}
\includegraphics[width=0.47\linewidth]{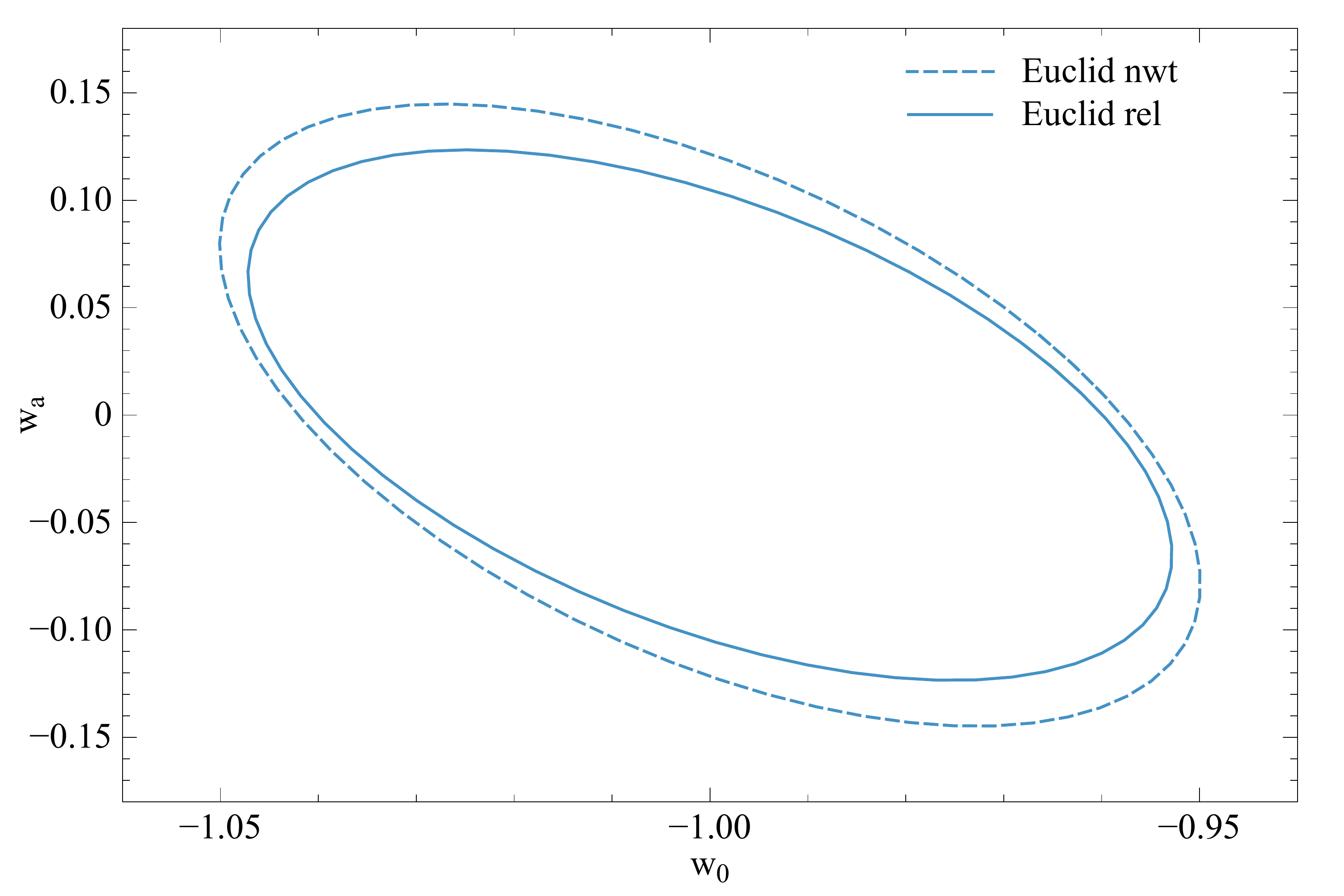}
\caption{
Constraints on dynamical dark energy parameters $\{w_0,w_a\}$ from SKA (spectroscopic, left) and Euclid (photometric, right). 
}
\label{fig:de}
\end{figure*}

In Figure~\ref{fig:de} we show the constraints on these dynamical dark energy parameters for the SKA and Euclid, marginalized over other cosmological parameters (here and throughout the entire paper we will show 1-$\sigma$ constraints).
It can be seen that, as expected, differences in the estimation of the errors are larger in the photometric case. We will quantify the error on the forecasted errors in the determination of dynamical dark energy parameters in Section~\ref{sec:results}.
For spectroscopic surveys, including lensing, and especially when magnification bias $s(z)$ is consistently included, actually degrades the precision with which $(w_0,w_a)$ can be determined by this observation. 

\subsection{Initial conditions from Inflation}
We study how large-scale correlations can be used to test the initial conditions of scalar perturbations. In particular, we focus on the spectral index, $n_s$ of the power spectrum and its running, $\alpha_s$ as well as the primordial non-Gaussianity parameter $f_{\rm NL}$ and its spectral tilt $n_{NG}$.

\subsubsection{The primordial  Power spectrum}
Inflation predicts a spectrum of primordial curvature and density perturbations  that are the  seeds of cosmic structure which forms as the universe expands and cools. 
This initial spectrum, and in particular its spectral index $n_s$ is a key observable for understanding inflation.
The spectrum is defined by:
\begin{equation}
k^3\langle {\mathcal R}(\bk) {\mathcal R}(\bk')\rangle = \de(\bk-\bk')P_{\mathcal R}(k) \,.
\end{equation}
The Dirac-delta is a consequence of statistical homogeneity and because of statistical isotropy $P_{\mathcal R}$ only depends on the modulus of $\bk$.
The spectral index $n_s$ is given by:
\begin{equation}
n_s - 1 \equiv \left.\frac{d \, {\rm ln} \, P_{\mathcal R}}{d \, {\rm ln} \, k}\right|_{k=k_*} \, ,
\end{equation}
where $k_*$ denotes the pivot scale which is usually chosen to be a scale to which a given experiment is most sensitive.
Given this, and to simplify the comparison with CMB experiments \cite{Ade:2015lrj}, we choose $k_*=0.05/$Mpc.
The quantity $n_s-1$ is also called the spectral tilt.
Inflation predicts a nearly scale-invariant spectrum of primordial density fluctuations, i.e. $n_s \approx 1$ and a nearly scale independent tilt, i.e. $n_s$ nearly independent of the pivot scale.
We characterize a possible scale dependence of  $n_s$ by:
\begin{equation}
\alpha_s = \left.\frac{d \, n_s}{d \, {\rm ln} \, (k)} \right|_{k=k_*}\, ;
\end{equation}
the quantity $\alpha_s$ is called the ``running'' of the spectral index.

\begin{figure*}[htb!]
\includegraphics[width=0.47\linewidth]{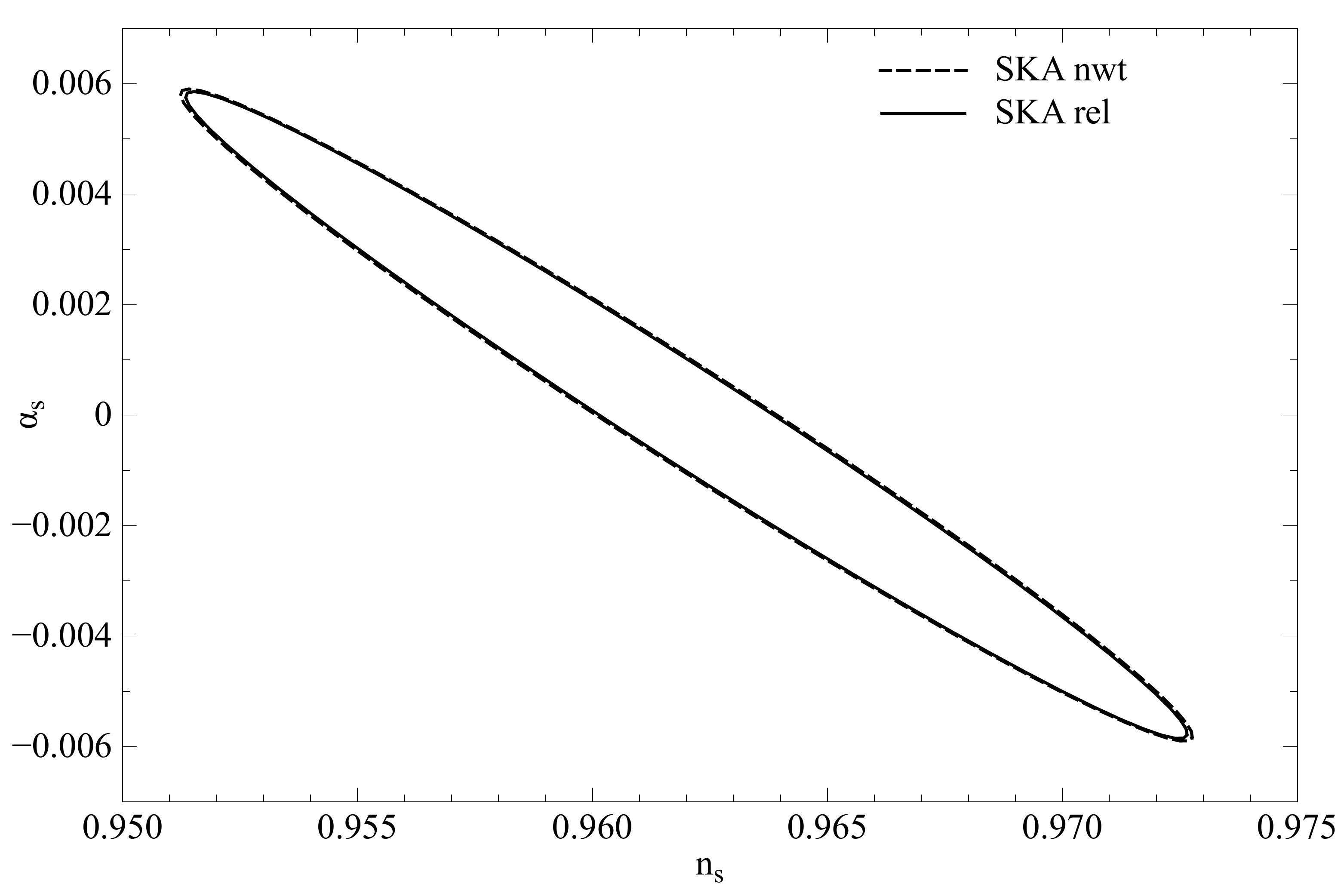}
\includegraphics[width=0.47\linewidth]{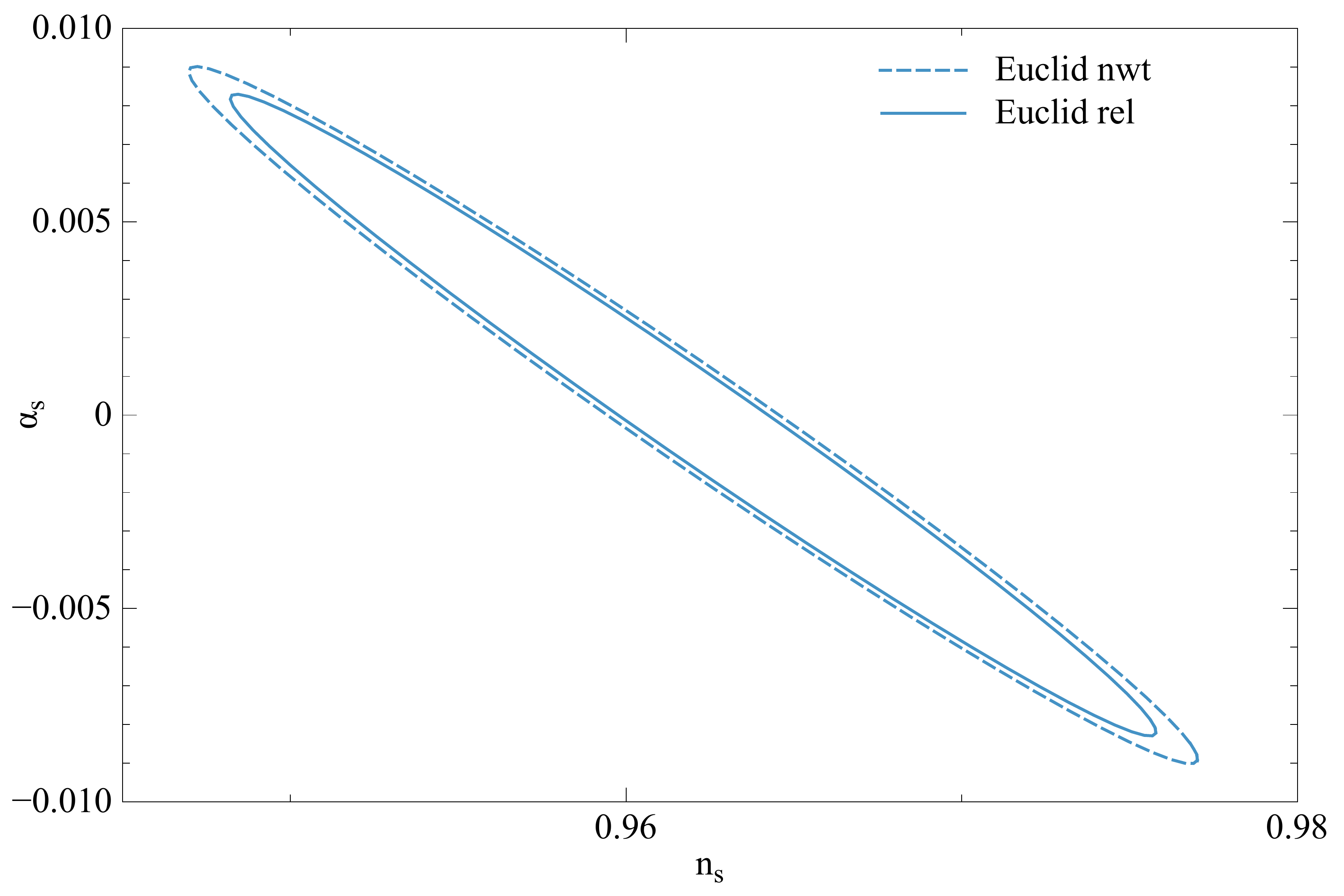}
\caption{
Comparison of the error ellipses for  the spectral index and its running in the Newtonian (dashed) and in the relativistic analyses (solid) for SKA (spectroscopic, left panel) and Euclid (photometric, right panel).
}
\label{fig:ns}
\end{figure*}

Figure~\ref{fig:ns} shows the constraints on power spectrum spectral index parameters, marginalized over the other cosmological parameters.
The effects of integrated terms is in this case smaller, and in the case of spectroscopic surveys, neglecting relativistic effects will induce only a $\sim 5\%$ error in the estimated precision, as we will see in Section~\ref{sec:results}.
It is however worth noting that the results here are obtained by marginalizing over the remaining cosmological parameters, while in Section~\ref{sec:results} we fix them.
In the photometric case, on the other hand, relativistic effects contribute significantly to the information content of galaxy clustering over the entire survey.

\subsubsection{Primordial non-Gaussianity}
An important goal for forthcoming cosmological experiments is to test whether initial conditions of cosmological perturbations deviate from Gaussianity. This is very interesting as standard slow roll single field inflation predicts very small non-Gaussanities~\cite{Maldacena:2002vr}.
This can be tested using CMB data (\cite{Bartolo:2004, Komatsu:2010, Ade:2015ava} and references therein) or using galaxy catalogs~\cite{Matarrese:2000, Dalal:2008, Slosar:2008, Matarrese:2008, Desjacques:2010, Xia:2010, Raccanelliradio, Ross:2013, Camera:2013, Ferramacho:2014, Raccanelli:2014ISW, Camera:2014, Camera:2014b, dePutter:2014, Alvarez:2014, SKA:Camera, Raccanelli:2015ani}. Constraining primordial non-Gaussianity offers a powerful test of the generation mechanism of cosmological perturbations in the early universe. While standard single-field models of slow-roll inflation lead to small departures from Gaussianity, non-standard scenarios (such as e.g. multi-field inflation) allow for a larger level of non-Gaussianity (see e.g.~\cite{Bartolo:2004, Komatsu:2010}).
A widely used parameterization of primordial non-Gaussianity is the $f_{\rm NL}$ parameterization, that includes a quadratic correction to the potential:
\begin{equation}
\label{eq:fnl}
\Phi_{\rm NG}=\phi+f_{\rm NL}\left(\phi^2-\langle\phi^2\rangle\right) ,
\end{equation}
where $\Phi_{\rm NG}$ denotes the Bardeen potential, which, on sub-Hubble scales reduces to the usual Newtonian gravitational potential.
Here $\phi$ is a Gaussian random field, and the second term, when $f_{\rm NL}$ is not zero, gives the deviation from Gaussianity.
In this paper we refer to the so-called ``local type" $f_{\rm NL}$. We use the LSS convention (as opposed to the CMB one, where $f_{\rm NL}^{\rm LSS} \approx 1.3 f_{\rm NL}^{\rm CMB}$ see \cite{Xia:2010} and Appendix~\ref{sec:fNL_class}).

One method for constraining non-Gaussianity from large-scale structure surveys exploits the fact that a positive $f_{\rm NL}$ corresponds to positive skewness of the density probability distribution, and hence an increased number of massive objects~\cite{Matarrese:2000, Dalal:2008, Desjacques:2010}.
In particular, $f_{\rm NL}$ introduces a scale-dependent modification of the large-scale halo bias. The difference from the usual Gaussian bias is given by~\cite{Dalal:2008}:
\begin{equation}
\label{eq:ng-bias}
\Delta b(z, k) = [b_{\rm G}(z)-1] f_{\rm NL}(k)\delta_{\rm ec} \frac{3 \Omega_{0m}H_0^2}{c^2k^2T(k)D(z)}, 
\end{equation}
where $b_{\rm G}(z)$ is the usual bias calculated assuming Gaussian initial conditions, which we assume to be scale-independent, $D(z)$ is the linear growth factor, $\Omega_{0m}$ is the matter density parameter today and $\delta_{\rm ec}$ is the critical value of the matter overdensity for ellipsoidal collapse. We choose to approximate it by the spherical collapse value $\delta_{\rm sc} \approx 1.68$ \cite{Coles:1995bd}.

While generally the parameter $f_{\rm NL}$ is assumed to be scale-independent, some inflationary models predicts a scale-dependent $f_{\rm NL}$, see e.g.~\cite{Chen:2005, Liguori:2006, Khoury:2009, Byrnes:2010, Riotto:2011, Kobayashi:2012} for more details.
The running parameter \nngt is expected to be $\lesssim\mathcal{O}(1)$ in most inflationary models. Its signatures in the CMB and LSS have been discussed in the literature, see e.g.~\cite{Loverde:2008, Sefusatti:2009, Becker:2011, Becker:2012, desjacques1, desjacques2}.
We consider the scenario that isolates the multi-field effects as in~\cite{Shandera, RaccanellifNL}, for which the effective $f_{\rm NL}(k)$ becomes:
\begin{equation}
\label{eq:biasfNL}
f_{\rm NL}(k) = f_{\rm NL} \left(\frac{k}{k_{*,\rm{NG}}}\right)^{n_{\rm NG}} \, ,
\end{equation}
where we choose the pivot scale $k_{*,\rm{NG}}=0.04/$Mpc to easily compare analyses of the CMB \cite{LoVerde:2007ri}. Effects of primordial non-Gaussianity on the bias are relevant on large scales due to the factor $k^{-2}$ like the relativistic effects~\cite{Maartens:2012}, so we may encounter a certain degree of degeneracy. However, as
 their angular dependence is different, they can in principle be distinguished in a multipole expansion~\cite{Raccanelli3D}.

\begin{figure*}[htb!]
\includegraphics[width=0.49\linewidth]{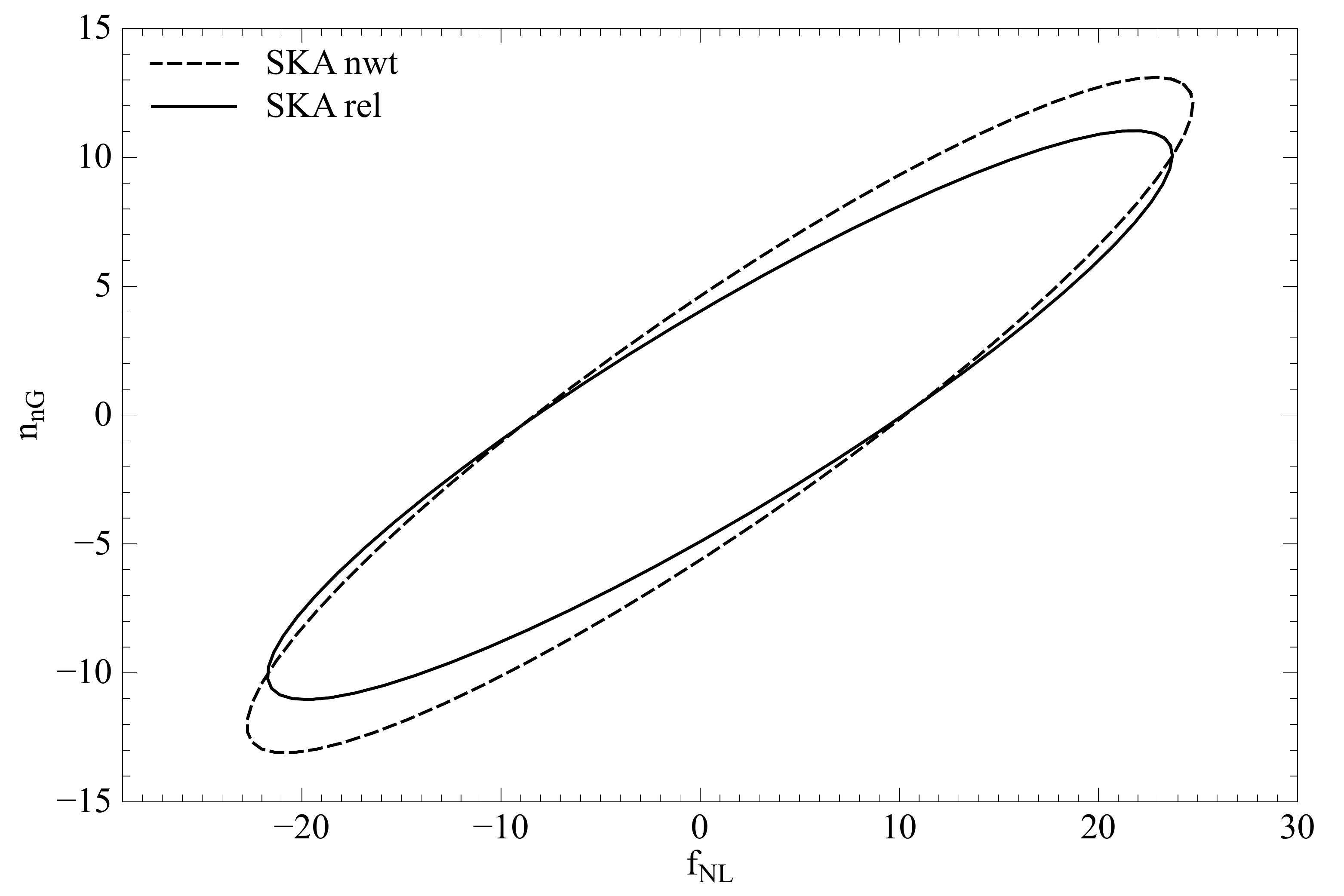}
\includegraphics[width=0.49\linewidth]{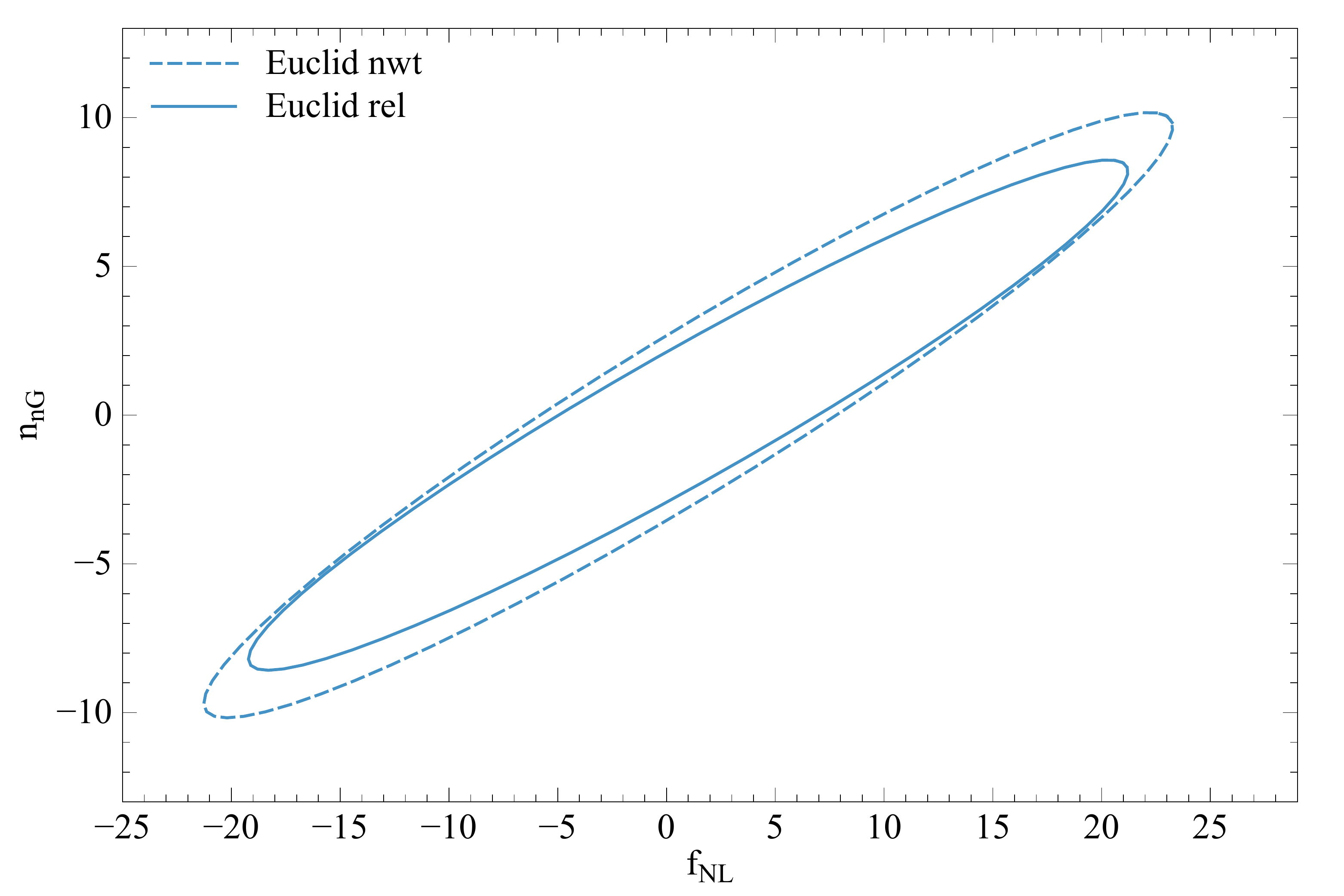}
\caption{
The same as Figs.~\ref{fig:de} and \ref{fig:ns}, for primordial non-Gaussianity parameters.
}
\label{fig:ng}
\end{figure*}

Figure~\ref{fig:ng} shows the constraints on primordial non-Gaussianity parameters, marginalized over other cosmological parameters.
We can see that in this case neglecting large-scale effects has a large impact, confirming, with a different analysis, the findings of~\cite{Camera:2014, Camera:2014b}.
We expect that in general lensing convergence is the dominant term in the radial cross-bin correlations (see~\cite{Raccanelliradial, DiDio:2013}), but since lensing is a tracer of the matter distribution and insensitive to galaxy bias, it is usually not taken into account when estimating non-Gaussianity from LSS. However, cosmic magnification effects, as from see Equation~\ref{eq:cosmag}, modify the effect of $f_{\rm NL}$ on clustering.

The variation in the ellipses is again larger in the photometric case, but it is still of $\sim 20\%$ for spectroscopic surveys.
We stress again that our results here are not a prediction of the final constraining power of future surveys. Several differences between our analysis and the ones of e.g.~\cite{Ferramacho:2014, Camera:2014} give us different predicted constraints for the SKA (e.g. the number of bins, number of parameters over which we marginalize, inclusion of running of $f_{\rm NL}$).
Given the precision in the measurements of non-Gaussianity parameters that is forecasted for future galaxy surveys when performing an optimized analysis~\cite{Ferramacho:2014, spherex, dePutter:2014}, a $\sim 20\%$ effect on the estimate of the error can be important for drawing theoretical conclusions from observations (see e.g.~\cite{Alvarez:2014}).

It is also worth noting that, as can be seen from Figure~\ref{fig:ng}, including relativistic effects changes the degeneracy between $f_{\rm NL}$ and $n_{\rm NG}$.

\section{Cosmological information in relativistic corrections}
\label{sec:cosdep}
In this Section we investigate the cosmological information present in the relativistic corrections, given by the lensing and gravitational terms of Equation~(\ref{eq:deltas}).

As stated in Section~\ref{sec:correlations}, lensing and gravitational potential contributions modify the observed correlations, acting as corrections to the intrinsic clustering. Here we study the information contained in these relativistic corrections, by isolating them and performing a Fisher analysis using signal coming only from these effects (i.e. excluding galaxy clustering signal).

\begin{figure*}[htb!]
\includegraphics[width=0.49\linewidth]{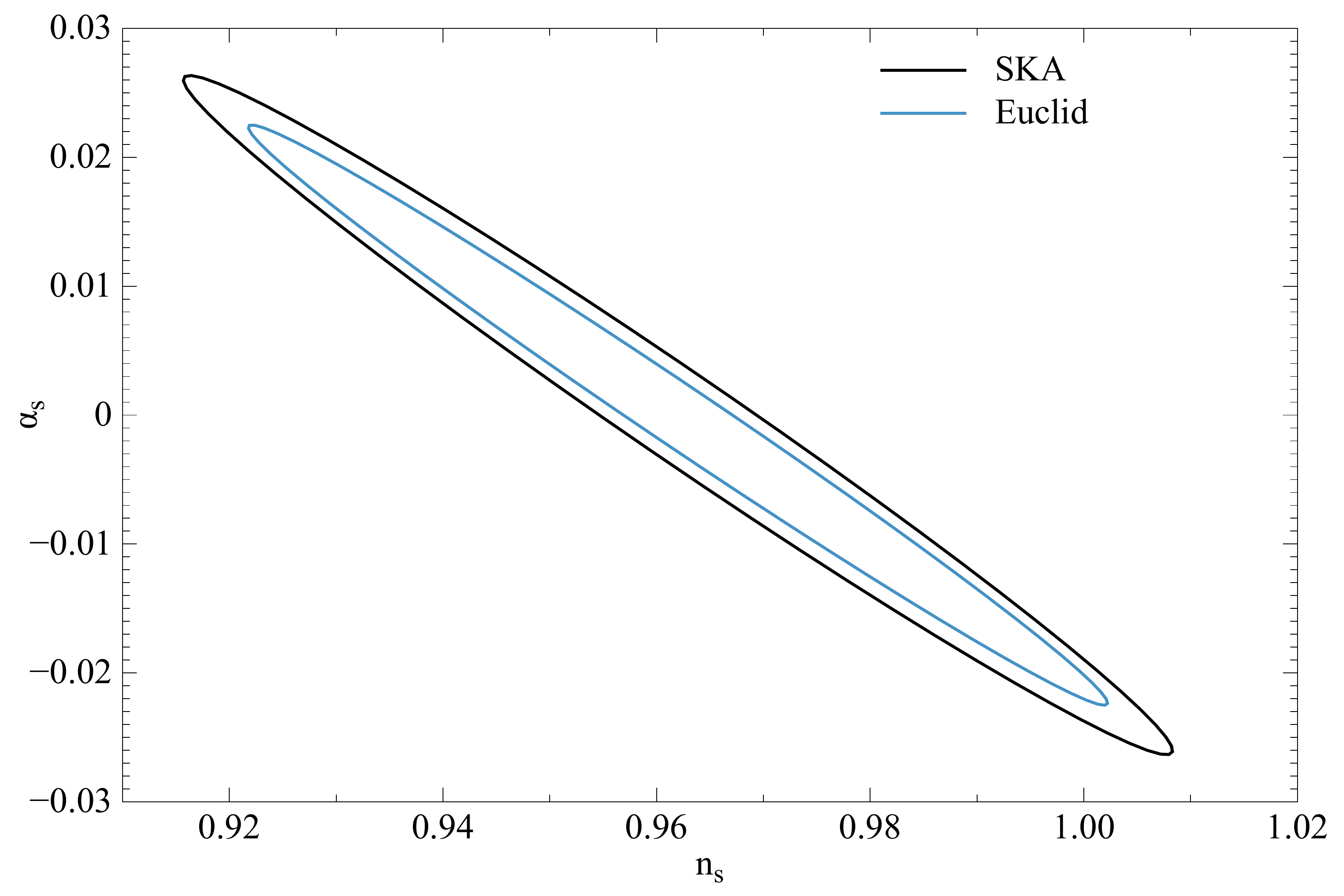}
\includegraphics[width=0.49\linewidth]{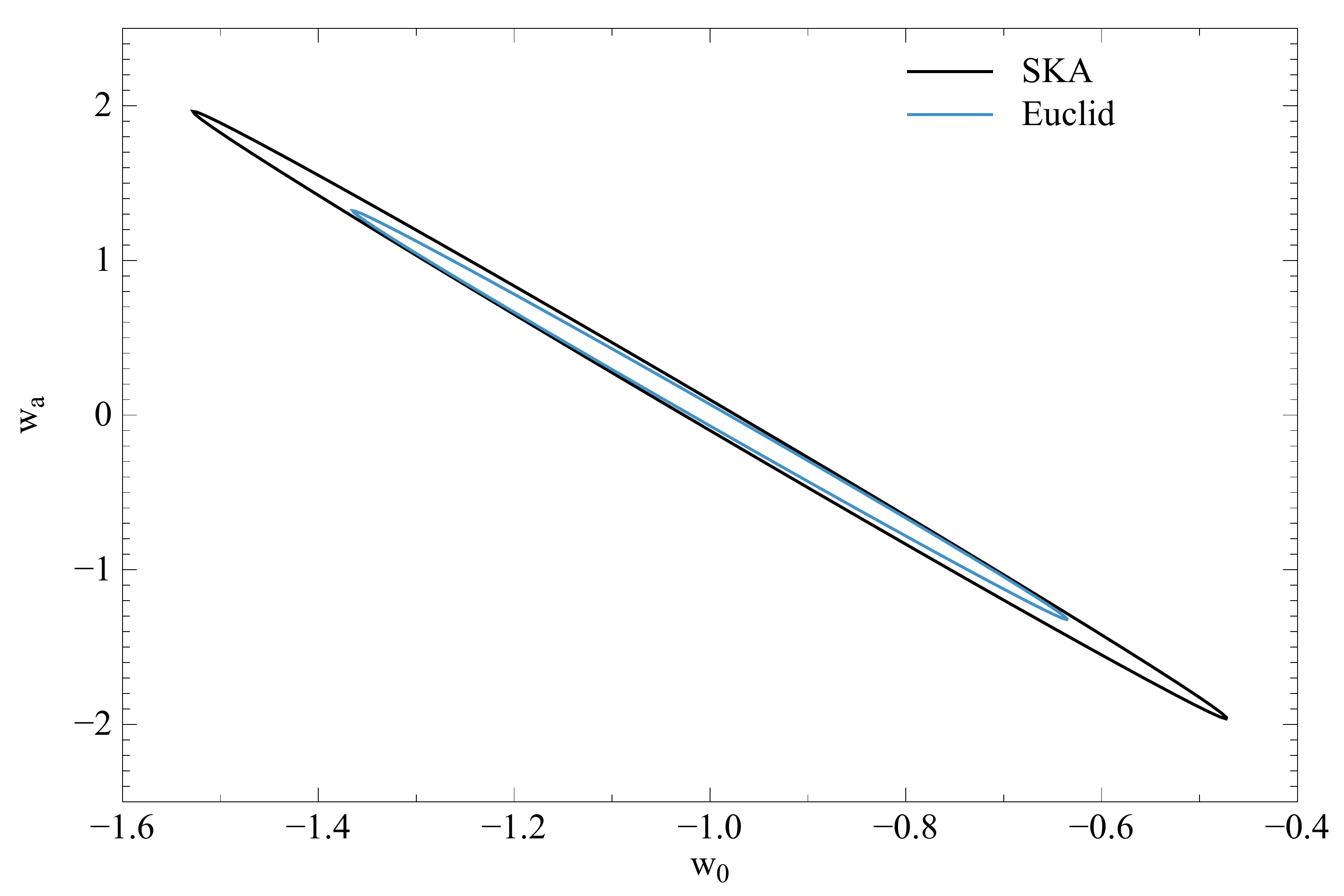}
\caption{
Constraints on $\{n_s, \alpha_s \}, \{w_0, w_a \}$ from the integrated terms in Equation~(\ref{eq:deltas}) for SKA (black) and Euclid (blue). 
}
\label{fig:cosdep_lengr}
\end{figure*}

In Figure~\ref{fig:cosdep_lengr} we show the constraints on cosmological parameters obtained by isolating the relativistic contributions. We marginalize over the cosmological parameters not considered in the given ellipse.
Clearly, these terms depend on cosmological parameters, so they can in principle be used for testing cosmological models, if they can be isolated; but more importantly, neglecting them in an analysis causes a bias of the results.

The constraining power of the integrated terms is of course smaller than the analysis that include galaxy clustering, but we can see how there clearly is some cosmological information in these correlations, and this explains the change it the ellipses in Section~\ref{sec:cosmology}.

It is interesting to note the nearly perfect degeneracy in the $\{w_0, w_a\}$ plane. Lensing is very sensitive to the combination $w_0+\frac{4}{15}w_a$, which it constrains to be $-1 \pm 0.03$.

These results also suggest that futuristic surveys or improved analyses (e.g. using the so called multi-tracer technique, or specifically planned observations targeting galaxy samples with a very large magnification bias) could provide a complementary test for cosmological models. In particular, integrated terms may contain additional information on the model of gravity; an analysis of this for the $f(R)$ model can be found in~\cite{Lombriser:2013}, and we will present a general investigation in a follow-up paper.

\section{Summary of results}
\label{sec:results}
\begin{figure*}[htb!]
\includegraphics[width=0.87\linewidth]{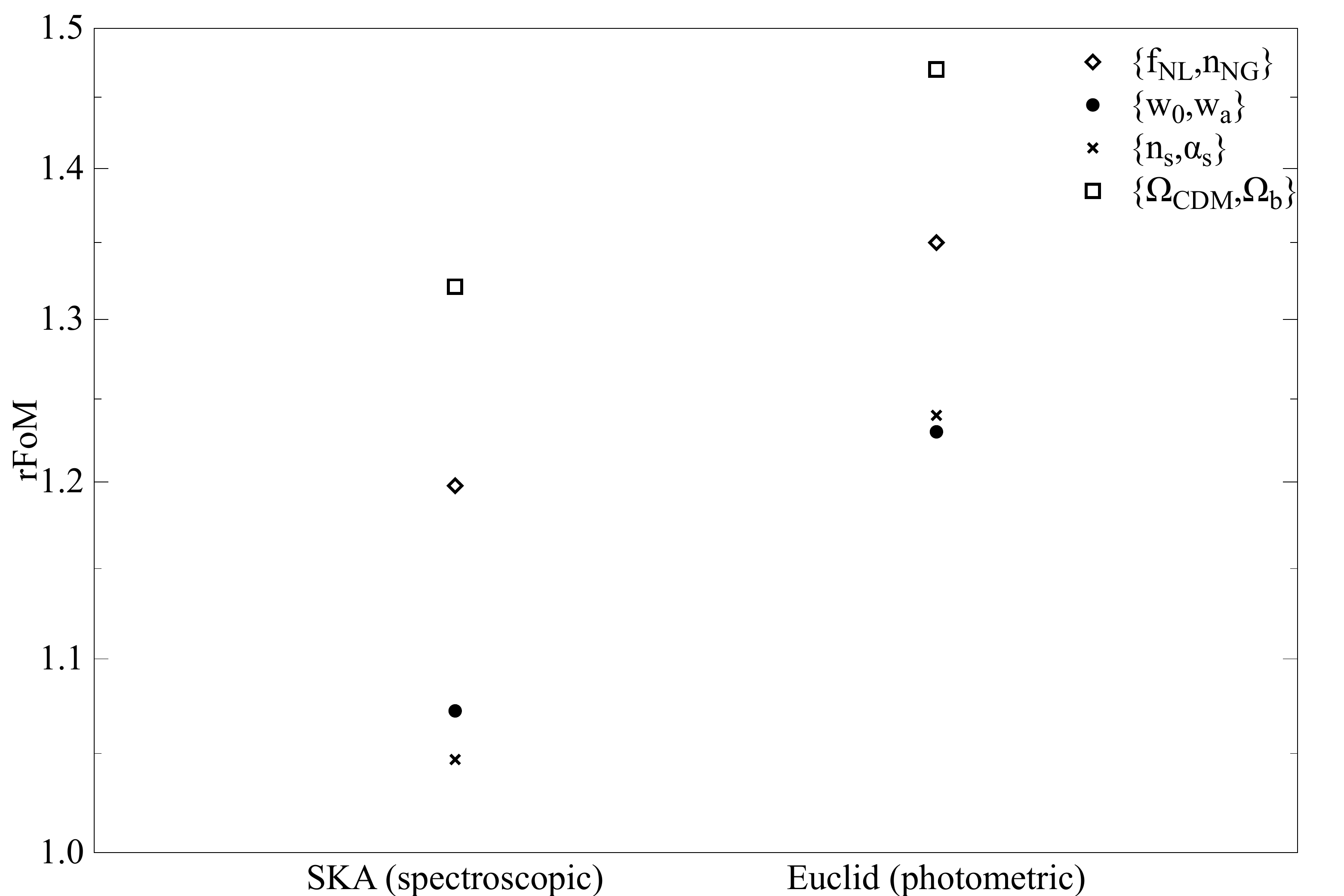}
\caption{
Modifications in the Figure of Merit for different sets of cosmological parameters when going from the Newtonian to the relativistic analysis, for the photometric Euclid and the spectroscopic SKA galaxy surveys.
}
\label{fig:FoM}
\end{figure*}

To quantify and summarize our results, we construct Figures of Merit (FoM) describing parameters of cosmological models. These are given by the inverse of the area of the error ellipses shown in the previous figures, see e.g.~\cite{DiDio:2013} for details. To properly quantify the error in the forecasted precision in measuring cosmological parameters introduced when neglecting relativistic effects, we then define the ratio:
\begin{equation}
\label{eq:FoM}
r{\rm FoM} = \frac{{\rm FoM}^{\rm rel }}{{\rm FoM}^{\rm nwt }} \, , \qquad {\rm FoM} = \left({\rm Det}  [(F_{\{A, B \}})^{-1}] \right) ^{-\frac{1}{2}} \, , 
\end{equation}
where $\{A, B \} = \{f_{\rm NL}, n_{\rm nG} \}, \{w_0, w_a \}, \{n_s, \alpha_s \}, \{\Omega_{\rm cdm}, \Omega_{\rm b} \}$ describe the model parameters we want to test, and $F_{\{A, B \}}$ is the sub-matrix of the Fisher matrix of the parameters $\{A,B \}$. This corresponds to fixing other parameters; clearly, the larger the FoM the better the given parameters are determined by the experiment.
We compute the FoM for the spectroscopic and photometric example surveys (SKA and Euclid, respectively). In Figure~\ref{fig:FoM} we show the errors in the in the predicted constraining power when performing a full relativistic analysis compared to the standard Newtonian analysis for all parameter subsets.

We can see that in most cases there is a significant difference in the FoM for cosmological models when using a full relativistic analysis rather than a standard Newtonian one. This shows that a precise calculation of the FoM for future surveys requires a proper relativistic formulation of galaxy clustering.
The largest effects are on the estimate of the matter content of the Universe and parameters describing primordial non-Gaussanity.
As described in the previous Sections, effects are larger for photometric surveys, due to the fact that photometric bins will cause an increased importance of the integrated radial terms coming from lensing and gravitational potential with respect to the intrinsic galaxy clustering signal.


\section{Conclusions}
\label{sec:conclusions}
In this paper we investigate the effect of including relativistic corrections to galaxy clustering on the constraining power of future galaxy surveys such as SKA and Euclid, by using the directly observable angular power spectra.
We study how cosmological parameter estimation is affected by relativistic terms by comparing the fully relativistic result with a standard Newtonian analysis.

We also show that cosmic magnification together with other relativistic terms, via their effective ``redistribution'' of galaxies in redshift space, introduce a bias on the constraints on cosmological, including non-Gaussianity parameters, and so it will need to be taken into account when forecasting measurements of e.g. $f_{\rm NL}$.

We analyze separately the integrated relativistic terms to determine their individual contributions to the signal and we show that they contain cosmological information.
While the constraining power of the relativistic terms is smaller than the one of the standard density and redshift space distortion terms, neglecting them leads to a considerable bias in the estimated precision in the measurements of cosmological parameters describing the models for dark energy and for the initial perturbations. We quantify our findings in Section~\ref{sec:results}, and we show that the error in neglecting relativistic terms is on the order of tens of percent, and of $\approx 20-40\%$ in most cases considered in this work.
Future surveys aiming to measure cosmological parameters with high precision and to set strict constraints on theoretical models will then need to take properly into account radial correlations and relativistic terms in order to not bias their results.

We don't perform a detailed analysis of the constraining power of the surveys, as that is beyond the scope of this work; we expect the FoM to improve with the number of bins, that we keep to 5, in order to speed up computations and to have an easier understanding of the different contributions to the final results.
Even though in slimmer bins the shot noise is increased due to the smaller number of galaxies, also the signal is increased since there is less ``smearing out'' of the dominant  fluctuations on smaller scales so that the noise increase can be compensated (see~\cite{DiDio:2013} for details).
It is an interesting future project to see how much can be gained by significantly increasing the number of bins.

Finally we want to note that the angular power spectrum approach of projecting sources on a single value of z within the redshift bin misses the radial information in it.

\vspace{0.5 cm}

{\bf Acknowledgments:}\\
We thank Roy Maartens for useful contributions and for reading the manuscript, and David Alonso, Stefano Camera, Enea Di Dio and Julien Lesgourgues for helpful discussions.\\
AR is supported by the John Templeton Foundation.
Part of the research described in this paper was carried out at the Jet Propulsion Laboratory, California Institute of Technology, under a contract with the National Aeronautics and Space Administration.
FM and RD acknowledge financial support by the Swiss National Science Foundation.
During the preparation of this work DB was supported by the Deutsche Forschungsgemeinschaft through the Transregio 33, ÔThe Dark UniverseÕ, and by the South African Square Kilometre Array Project. 
FM and DB aknowledge the hospitality of the Department of Physics \& Astronomy, Johns Hopkins University, where this work was completed.

\vspace{2.5 cm}

\appendix
\section{Implementation of primordial non-Gaussianity effects in {\sc Class}}
\label{sec:fNL_class}
In this Appendix we summarize our modifications to the {\sc Class} code \cite{Blas:2011rf,DiDio:2013bqa} to include effects of primordial non-Gaussianity effects, and show the number counts transfer functions computed in the code.
Before we enter into this we want to clarify the definitions given at the beginning of the paper.
\subsection{Definitions and notation}
The relativistic expressions for the number counts in linear perturbation theory is given in Equation~(\ref{eq:deltas}),
\begin{equation}
\label{eqA:deltas}
\Delta_{\rm obs} ({\bf n}, z) = \Delta_{\delta}({\bf n}, z) + \Delta_{\rm rsd}({\bf n}, z) + \Delta_{\rm v}({\bf n}, z) + \Delta_{\rm \kappa}({\bf n}, z) + \Delta_{\rm pot}({\bf n}, z) \, ,
\end{equation}
where~\cite{Bonvin:2011,Challinor:2011}
\begin{eqnarray}
\Delta_{\delta}({\bf n}, z)  &=& b(z,k)\;\de_{\rm co}\left(r(z)\bn,\tau(z)\right) \\
\Delta_{\rm rsd}({\bf n}, z)  &=& \frac{1}{\HH(z)}\dd_r(\bV\cdot\bn) \\
\Delta_{\rm v}({\bf n}, z)  &=& \left[\frac{\HH'}{\HH^2}+\frac{2-5s(z)}{r\HH} + 5s(z)- f_{\rm evo}(z)\right](\bV\cdot\bn)+ \left[3\HH-f_{\rm evo}(z)\right]\De^{-1} (\nabla\cdot\bV) \\
\Delta_{\rm \kappa}({\bf n}, z)  &=&(2-5s(z))\ka =-\frac{(2-5s(z))}{2}\int_0^{r(z)}dr\frac{r(z)-r}{r(z)r}\De_2(\Phi+\Psi) \\
\Delta_{\rm pot}({\bf n}, z)  &=&  (5s(z)-2)\Phi+\Psi +\HH^{-1}\Phi' +  \left[\frac{\HH'}{\HH^2}+\frac{2-5s}{r\HH}+ 5s(z)- f_{\rm evo}(z)\right]\left[\Psi +\int_0^{r(z)}dr(\Phi'+\Psi')\right] +\nonumber \\
&& \quad + \frac{2-5s}{r(z)}\int_0^{r(z)}dr(\Phi+\Psi) \,.
\end{eqnarray}
Here $\bV$ is the peculiar velocity, $\Phi$ and $\Psi$ are the Bardeen potentials, $\de_{\rm co}$ is the density contrast in comoving gauge and $\HH = aH$ is the conformal Hubble parameter.   All quantities which are not integrated are evaluated at conformal time $\tau(z)$ and at position $r(z)\bn=(\tau_0-\tau(z))\bn$. Here $r(z)$ is the conformal distance on the light cone, $r(z)=\tau_0-\tau(z)$. A prime indicates a derivative w.r.t. conformal time. The  term $\De^{-1}(\nabla\cdot\bV)$ is the velocity potential. It comes from the {\it non-local} requirement that the overdensity $\de_{\rm co}$ which is multiplied with the bias factor $b(z)$ be the comoving one. In $k-$space this becomes simply $-V(k)/k$.

Expanding these expressions in spherical harmonics and defining by $S_X(k,\tau)$ the solution of linearized Einstein equations for the variable $X$ with initial condition $\Phi(k,\tau_{\rm in})=\Psi(k,\tau_{\rm in})=1$ and $\Phi'(k,\tau_{\rm in})=\Psi'(k,\tau_{\rm in})=0$, where $\tau_{\rm in}$ is such that the corresponding mode is super horizon, i.e. $k\tau_{\rm in}\ll 1$, we find after integration over a normalized window function $W_i(z)$ centered at $z_i$ that the transfer function $\Delta_{\ell}^{W_i}$ given in Equation~(\ref{eq:Cls}) is the sum of the following terms (see~\cite{DiDio:2013bqa} for more details):
\begin{eqnarray}
\Delta_{\ell}^{\mathrm{Den}_i} &=& \int_0^{\tau_0} d\tau W_i \, \left[ b_G(z)+\Delta b(z,k) \right] S_\mathrm{\delta} \, j_{\ell} \nonumber \\
\Delta_{\ell}^{\mathrm{Len}_i} &=& \ell(\ell+1) \int_0^{\tau_0} d\tau \, W^\mathrm{L}_i \,  S_{\Phi+\Psi} \, j_{\ell} \nonumber \\
\Delta_{\ell}^{\mathrm{V}1_i} &=& \int_0^{\tau_0} d\tau \, W_i \left[ 1 \! + \! \frac{H'}{aH^2} \! + \frac{2 -5s }{(\tau_0-\tau)aH }  \!  +5s - f_{\rm evo}  \right] \frac{S_{\Theta}}{k} \, j'_{\ell}  \nonumber \\
\Delta_{\ell}^{\mathrm{V}2_i} &=& \int_0^{\tau_0} d\tau \, W_i \left(f_{\rm evo} -3 aH\right) \frac{S_{\Theta}}{k^2} \, j_{\ell} \nonumber \\
\Delta_{\ell}^{\mathrm{V}3_i} &=& \int_0^{\tau_0} d\tau \, W_i \left( \frac{1}{aH} \right) S_{\Theta} \, j''_{\ell} \nonumber \\
\Delta_{\ell}^{\mathrm{G}1_i} &=& \int_0^{\tau_0} d\tau \, W_i \, S_\Psi \, j_{\ell} \nonumber \\
\Delta_{\ell}^{\mathrm{G}2_i} &=& -\int_0^{\tau_0} d\tau \, W_i  \left[3+\frac{H'}{aH^2} + \frac{2 -5s }{(\tau_0-\tau)aH } -f_{\rm evo}  \right] S_\Phi \, j_{\ell} \nonumber \\
\Delta_{\ell}^{\mathrm{G}3_i} &=& \int_0^{\tau_0} d\tau \, W_i \, \left( \frac{1}{aH} \right) S_{\Phi'} \, j_{\ell} \nonumber \\
\Delta_{\ell}^{\mathrm{G}4_i} &=& \int_0^{\tau_0} d\tau \, W_i^{\mathrm{G}4} \,  S_{\Phi+\Psi} \, j_{\ell} \nonumber \\
\Delta_{\ell}^{\mathrm{G}5_i} &=& \int_0^{\tau_0} d\tau \, W_i^{\mathrm{G}5} \, S_{(\Phi+\Psi)} k\, j'_{\ell} ~.
\label{eq:delta_terms}
\end{eqnarray}
Here we have also allowed for a scale dependent bias $\De b(z,k)$ which can be added to the Gaussian bias $b_G(z)$. We shall introduce this in the next subsection.
Furthermore, also the functions $s(z)$ and $f_{\rm evo}(z)$ depend on redshift as described in the main text.

As in Ref.~\cite{DiDio:2013bqa}, the different contributions correspond to the density in comoving gauge $\delta_{\rm co}$ (``Den''), lensing convergence $\kappa$ (``Len''), Doppler (``V1''-``V2''), redshift-space distortions in the Kaiser approximation (``V3'') and terms depending on gravitational potential (``G1''-``G5''), respectively.
We use conformal time $\tau$, the proper Hubble parameter $H$ and the scale factor $a$.
We have omitted the arguments $k$ for the transfer functions $\Delta$, $(\tau,k)$ for the source functions $S_X$, $rk$ for the Bessel functions $j_{\ell}$, and $\tau$ for selection and background functions.
Note that, for consistency with the {\sc Class} code, the velocity source function $S_\Theta(\tau,k)$ is given by $\Theta(k) \equiv kV(k)$, where $V(k)$ is the velocity perturbation in the Newtonian gauge.
Primes indicate derivatives with respect to the argument.
The index $i$ refers to the redshift bin around reference redshift $z_i$.
For the integrated terms ``Len'', ``G4'' and ``G5'', we have introduced
\begin{eqnarray}
W_i^\mathrm{L}(\tau) &=& \int_0^\tau \!\! d\tilde{\tau} W_i(\tilde{\tau}) \left( \frac{2-5s}{2} \right) \frac{\tau-\tilde\tau}{(\tau_0-\tau)(\tau_0-\tilde\tau)} \nonumber \\
W_i^{\mathrm{G}4}(\tau) &=&  \int_0^\tau \!\! d\tilde{\tau} W_i(\tilde{\tau})  k\,  \frac{2 -5s}{\tau_0-\tilde\tau} \\
W_i^{\mathrm{G}5}(\tau) &=& \int_0^\tau \!\! d\tilde{\tau} W_i(\tilde{\tau}) \left[1+\frac{H'}{aH^2} +  \frac{2 -5s}{(\tau_0-\tilde\tau) aH} + 5s-f_{\rm evo} \right]_{\tilde{\tau}}~. \nonumber
\end{eqnarray}
The window function $W_i(\tau)$ is given by the product of the number of galaxies per solid angle and redshift $dN/dz/d\Omega$  multiplied, in the spectroscopic case, by a tophat centered around $z_i$ and of the width of the bin and, in the photometric case, by a Gaussian function with standard deviation equal to the half-width of the bin (see e.g.~\cite{Blake:2005}).
For numerical convenience we avoided to express $S_{\Phi'}$ in terms of the time derivative $\Phi'$ that requires numerical derivatives of Einstein equations, instead $\Phi'$ can be obtained analytically from Einstein equations~\cite{Ma:1995ey}.
The corresponding expressions presented in \cite{DiDio:2013bqa} are recovered integrating by parts $\Delta_{\ell}^{\mathrm{G}5_i}$ (neglecting boundary terms since they vanish as $\tau\to0$ and are unobservable for $\tau=\tau_0$) and redefining consistently $\Delta_{\ell}^{\mathrm{G}1_i}$ and $\Delta_{\ell}^{\mathrm{G}2_i}$ so that the sum of these transfer functions coincides with the form given in previous work and in {\sc ClassGal}\footnote{\url{http://cosmology.unige.ch/content/classgal}}.
Note that the ``G4'' term contains the Shapiro time delay, while ``G5'' is related to the integrated Sachs-Wolf effect (using integration by parts, the derivative acting on the Bessel function can be interchanged with the time derivative of the sum of Bardeen potentials).

\subsection{Modifications of {\sc class}}
We now describe how to implement the $f_{\rm NL}$ parameter (and its running $n_{\rm NG}$) for local primordial non-Gaussianity (PNG).
In {\sc Class}, it is convenient to use the Poisson equation $\Phi=-\left( 3\Omega_{0 {\rm m}} H_0^2/ 2ak^2 \right) \delta_{\rm co}$ in Equation~(\ref{eq:ng-bias}), to obtain:
\begin{equation} \label{eq:Db_class}
\Delta b(z,k) =  -2(b_{\rm G}(z)-1) f_{\rm NL} \left( \frac{k}{k_*} \right)^{n_{\rm NG}} \delta_{\rm ec} \frac{g(0)}{g(z_{\rm dec})} \frac{2\mathcal R}{3\delta} \; ;
\end{equation}
for more details about the derivation of this result see~\cite{Dalal:2008}.
Here $f_{\rm NL}$, $n_{\rm NG}$, $k_*$ and $\delta_{\rm ec}$ are constants describing the PNG parameter, its running, its pivot scale and the critical density for ellipsoidal collapse, respectively.
The linear and scale-independent galaxy bias in absence of PNG is given by the function $b_{\rm G}(z)$.
The variable $\mathcal{R}=-3\Phi_p/2$ (here $\Phi_p$ is the primordial Bardeen potential), is the primordial potential of curvature perturbations.
The linear growth factor $g(z)=(1+z)D(z)$, where $D(z)$ is the amplitude of the growing mode, appears since {\sc Class}, by normalizing all source functions with respect to $\mathcal{R}$ (e.g., $\delta(z,k)/\mathcal{R}(k)$), would otherwise evaluate $\Phi$ of Equation~(\ref{eq:fnl}) at decoupling, while in LSS convention it is evaluated at $z=0$.
In other words, we take into consideration the following relation between LSS and CMB conventions~\cite{Xia:2010, Camera:2014}:
\begin{equation}
f_{\rm NL}^{\rm LSS} = \frac{g(z_{\rm dec})}{g(0)} f_{\rm NL}^{\rm CMB} \;,
\end{equation}
and we use $f_{\rm NL} \equiv f_{\rm NL}^{\rm LSS}$.

We refer to {\sc Class v2.3.2}\footnote{\url{http://class-code.net/}}, where a constant galaxy bias is implemented in the \texttt{transfer.c} module as a rescaling of the density transfer function.
As a first step, we remove this constant $b_{\rm G}$ from the \texttt{transfer.c} module.
As described in~\cite{DiDio:2013bqa}, the code computes the density perturbation in the comoving gauge, related to the longitudinal Newtonian gauge by:
\begin{equation}
\delta_{\rm co} = \delta_{\rm long}+3 \frac{aH}{k^2} \theta_{\rm long} \; ,
\end{equation}
where $\theta_{\rm long}=kV_{\rm long}$ is the divergence of the velocity in the longitudinal gauge.
Furthermore, $\delta_{\rm co}$ is a gauge-invariant variable and at first order in perturbation theory it coincides with the density perturbation in the synchronous-gauge comoving with dark matter used in~\cite{Dalal:2008}.
Then, we modify the \texttt{perturbations.c} module to rescale the matter density source function $\delta_{\rm co}(z,k)/\mathcal{R}(k)$ by a function $b_{\rm G}(z)$ specified within the same module.
Finally, we add the quantity $\Delta b\cdot\delta_{\rm co}/\mathcal{R}$ from Equation~(\ref{eq:Db_class}) to the rescaled density contrast $b_{\rm G}(z)\delta_{\rm co}/\mathcal{R}$.

With this modifications, and allowing also for a time-dependent magnification bias $s(z)$, the transfer functions given in Equation~(\ref{eq:Cls}) become the terms given in Equation~(\ref{eq:delta_terms}).

\end{document}